\documentclass[journal]{IEEEtran}

\usepackage{graphicx}

\usepackage[cmex10]{amsmath}
\usepackage{mathtools,amsthm}
\usepackage{times}
\usepackage{multicol}
\usepackage{amssymb, amsfonts}
\usepackage{cases}

\usepackage{enumerate}

\newcommand{\id}[1][3]{{I}}
\newcommand{\zero}[2]{{0}}
\newcommand{\zeros}[2]{{0}}

\newcommand{\multeqi}[2]{\begin{IEEEeqnarraybox}[][#2]{#1}}
\newcommand{\multeqf}{\end{IEEEeqnarraybox}}

\newcommand{\systemi}[1][rCL]{\left\lbrace\begin{IEEEeqnarraybox}[][c]{#1}}
\newcommand{\systemf}{\end{IEEEeqnarraybox}\right.}

\newcommand{\eqni}[1][rCL]{\begin{IEEEeqnarray}{#1}}
\newcommand{\eqnf}{\end{IEEEeqnarray}}

\newcommand{\nneqni}[1][rCL]{\begin{IEEEeqnarray*}{#1}}
\newcommand{\nneqnf}{\end{IEEEeqnarray*}}

\newcommand{\pmatrixi}{\begin{pmatrix}}
\newcommand{\pmatrixf}{\end{pmatrix}}

\newcommand{\bmatrixi}{\begin{bmatrix}}
\newcommand{\bmatrixf}{\end{bmatrix}}

\newcommand{\smatrixi}{\left[\begin{smallmatrix}}
\newcommand{\smatrixf}{\end{smallmatrix}\right]}

\newcommand{\enumi}{\begin{enumerate}}
\newcommand{\enumf}{\end{enumerate}}

\newcommand{\enumri}{\begin{enumerate}\renewcommand{\theenumi}{\textit{\roman{enumi}}}}
\newcommand{\enumrf}{\end{enumerate}}

\newcommand{\mytheorem}[2]{%
\newtheorem{t#2}{#1}%
\newenvironment{#2}{\begin{t#2}}{\end{t#2}}}

\theoremstyle{plain}
\newenvironment{remark}[1][Remark]{\begin{trivlist}
\item[\hskip \labelsep {\bfseries #1}]}{\end{trivlist}}

\mytheorem{Theorem}{theorem}
\mytheorem{Proposition}{proposition}
\mytheorem{Lemma}{lemma}
\mytheorem{Corollary}{corollary}
\mytheorem{Assumption}{assumption}
\mytheorem{Definition}{definition}
\mytheorem{Property}{property}

\newenvironment{system}[1][rCL]{\left\lbrace\begin{IEEEeqnarraybox}[][c]{#1}}{\end{IEEEeqnarraybox}\right.}

\usepackage{paralist}
\usepackage{textcomp}
\usepackage{gensymb}
\usepackage{color}

\graphicspath{figures/}

\newtheorem{hypothesis}{Assumption}

\usepackage[bookmarks=true]{hyperref}
\hypersetup{pdfinfo={
   Author={Daniele Pucci},
   Title={Analysis and Control of Aircraft Longitudinal Dynamics  with Large Flight Envelopes},
   CreationDate={D:20140625120000},
   Subject={Robotics},
   Keywords={Aerodynamic Force Modelling, Flight Equilibrium Analysis, Flight Control, Transition maneuvers},
}}

\begin{document}
\title{Analysis and Control of Aircraft Longitudinal Dynamics  
with Large Flight Envelopes}

\author{Daniele~Pucci
\thanks{The author is with the iCubFacility department, 
Istituto Italiano di Tecnologia, 16163 Genova,
Italy (e-mail: daniele.pucci@iit.it)}}

\markboth{Journal of \LaTeX\ Class Files,~Vol.~14, No.~8, August~2015}%
{Shell \MakeLowercase{\textit{et al.}}: Bare Demo of IEEEtran.cls for IEEE Journals}

\maketitle

\begin{abstract}
The paper contributes towards the development of a unified
control approach for longitudinal aircraft dynamics with large flight envelopes. Prior to the control design,
we analyze the existence and the uniqueness of the equilibrium orientation along a  reference velocity. 
We show that shape symmetries and aerodynamic stall phenomena imply the existence of the equilibrium orientation
irrespective of the reference velocity. The equilibrium orientation, however, is not in general unique, and this may 
trigger an aircraft loss-of-control for specific reference velocities.
Conditions that ensure the local and the global uniqueness of the equilibrium orientation are stated.
We show that the uniqueness of the equilibrium orientation is intimately related to the so-called
\emph{spherical equivalency}, i.e. the existence of a thrust change of variable
rendering the direction of
the transformed external  force 
independent of the vehicle's orientation, as in the case of spherical shapes.
Once this transformation is applied,
control laws for reference velocities can be designed. These laws extend the so-called \emph{vectored--thrust control paradigm} 
to the case of generic body shapes subject to steady aerodynamics.
\end{abstract}


\section{Introduction}

The profound human curiosity about nature's flight systems and the dream of flying 
have prompted the long, irregular, and faltering understanding of basic aerodynamic phenomena.
Yet, despite decades of research on aerodynamics and flight control,
 flying machines' dynamics are a far cry from being fully understood. 
This paper contributes towards the comprehension and the control of the aircraft
longitudinal dynamics 
 subject to steady aerodynamic forces.

Flight control makes extensive use of linear control techniques \cite{2003_STEVENS}. One reason why is the 
existence of numerous tools to assess the robustness properties of a linear feedback controller \cite{rdb12}. 
Another reason is that flight control techniques have been developed primarily for 
commercial airplanes, which are designed and optimized to fly along very specific trajectories. 
Control design is then typically achieved from the linearized equations of motion along desired trajectories that often represent
\emph{steady-state conditions}. 
Clearly, the hidden assumption behind linearization techniques is the existence of  
equilibrium conditions that --- to the best of the author's knowledge -- has never been investigated before.

Some aerial vehicles, however, are required to fly in very diverse conditions that involve
large variations of the angle of attack, i.e. \emph{large flight envelopes}. Fighter aircraft, convertible Vertical Take-Off and Landing (VTOL) aircraft, and small 
Unmanned Aerial vehicles (UAVs) in windy conditions are few examples of aircraft with large flight envelopes. 
It then matters to ensure large stability domains that are achievable via the use of nonlinear feedback designs. 

The nonlinear effects of large flight envelopes on aircraft dynamics are of potentially fatal
importance in practice. These effects, in fact, can give rise to an aircraft \emph{loss-of-control} (LOC), 
which remains one of the most important contributors to fatal accidents~\cite{2009_KWATNY,2010_BELCASTRO}. 
A number of different types of 
LOCs  in longitudinal and lateral/directional motion is related to \emph{stall phenomena}~\cite[s. 2.2]{1997_GOMAN}. Among these forms, one can 
mention a stable flight at high angle of attack without rotation, also called a \emph{deepstall} condition.
Other forms of LOCs are due to the \emph{roll-} and the \emph{inertia-coupled} problems 
\cite{1948_PHILLIPS,1974_HACKER,1998_JAHNKE}. 

Many types of aircraft LOCs are related to the 
\emph{equilibria pattern} variations 
depending on the vehicle control settings. 
In fact, aircraft dynamics may have more than one equilibrium point 
associated with a given control setting \cite[p. 728]{2004_STENGEL}: when this setting varies, 
the aircraft equilibrium  may \emph{jump} from one stable configuration to another, which  may  cause abrupt  aircraft responses  and, eventually, an aircraft LOC.  

The qualitative behavior of aircraft dynamics in relation to their  equilibria pattern
can be obtained from the \emph{bifurcation analysis and catastrophe theory methodology}~\cite{1982_CARROLL}.
For an introduction to bifurcation analysis of aircraft dynamics,
the reader is referred to \cite{1982_CARROLL,2004_CUMMINGS,2002_LOWENBERG}.
In essence, the  nonlinear problem 
is  formulated in the form of a set of ordinary differential equations depending on parameters \cite{1997_GOMAN}, which often 
represent the control surface deflections. Yet, the analysis of how the equilibria pattern varies versus the reference trajectory is still missing.  
We shall see in this paper that this analysis characterizes a set of non-trackable transition maneuvers between hovering and high velocity
cruising.

Once the bifurcations are identified, one may apply the so-called \emph{bifurcation control} to stabilize the system around the 
interested bifurcation. In particular, one can modify  the equilibria characteristics 
via a designed control 
input \cite{2000_CHEN}. Applications of
bifurcation control to aircraft dynamics are \cite{1991_KWATNY,1990_ABED}. 
The assumption that the system is autonomous, however,
clearly impairs the proposed control approach when the error dynamics is time dependent, as in the case of a time-varying reference. Also,
the effectiveness of the \emph{bifurcation control} is  related to the model of the chaotic region where bifurcations occur 
--~often the stall region~-- which is very difficult to model accurately.

Avoiding the conditions that may eventually trigger a LOC is then fundamental for any 
nonlinear feedback control.  Following  \cite{ss80}, control laws based on the 
dynamic inversion technique have been proposed to extend the flight envelope of military aircraft (see, e.g., \cite{ws05} and the 
references therein). The control design strongly relies on tabulated models of aerodynamic forces and moments, like the 
High-Incidence Research Model 
\cite{mu97}. 
Compared to linear techniques, this type of approach extends the flight domain without involving gain scheduling strategies, but the angle of attack is assumed to remain away from the
stall zone. However, should this assumption be violated the system's behavior is unpredictable. 

Compared to commercial airplane control, nonlinear  
control of VTOL vehicles is more recent, and it has been addressed with a larger variety of techniques, such as dynamic inversion \cite{hsm92}, Lyapunov-based design \cite{2002_MARCONI,2003_ISIDORI}, 
Backstepping \cite{Bouabdalla05}, Sliding modes \cite{Bouabdalla05,Xu08}, and Predictive control \cite{kim02,bph07-ifac} (see~\cite{hhms13} for more details). Most of these studies address the stabilization 
of hover flight or low-velocity trajectories: little attention has been paid
to aerodynamic effects, which are typically either ignored or modeled as additive perturbations. 

When considering the class of convertibles vehicles, one of the major control problems is related to the \emph{transition maneuvers} between
hovering and high-velocity cruising. During this transition, the aerodynamic effects become from negligible to preponderant,
and the issues related to the LOCs can be carefully dealt with.
Several studies have been dedicated to the control of  transition maneuvers for 
convertibles~\cite{2007_BENOSMAN,2007_FRANK,1999_OISHI,2010_DESBIENS,2011_NALDI,2012_CASAU}, the 
common denominator of which is a 
``switching'' policy between a \emph{hover} and a \emph{cruise control} depending on the actual flight state.  A technical difficulty is ensuring the stability of the closed-loop system along the transition, which is very sensitive to the ``switching'' policy  usually tuned for specific classes of reference trajectories. 
To the author's knowledge, a unified approach for the control of the transition maneuvers not based on a ``switching'' policy 
and allowing for large flight envelopes is  missing.

%

This paper extends and  encompasses the contributions~\cite{phms11CDC,2012_PUCCI1,2013_PUCCI}  on the longitudinal dynamics control of aerial vehicles subject to steady aerodynamic forces. 
Given a reference velocity, we first investigate the 
\emph{existence} of an equilibrium orientation about which the vehicle must be stabilized: the existence of an equilibrium 
orientation along any reference velocity follows from the symmetry of the vehicle's shape and/or aerodynamic stall phenomena. 
Also, for bi-symmetric shapes, we show that
there exists an equilibrium orientation that ensures a positive-thrust.
We then study the \emph{multiplicity} of the equilibrium orientation in the case of NACA airfoils. The main outcome of this
analysis is that a constant-height transition maneuver between hovering and high-velocity cruising cannot be in general perfectly tracked because of stall phenomena, which may trigger a LOC. 
For control design purposes, we  analyze the \emph{local} and \emph{global} uniqueness of the equilibrium 
orientation. Apart from pathological cases, when the equilibrium orientation is locally unique, we show that the system can be transformed
into a form where the equivalent external force has a direction independent of
the vehicle's orientation. This transformation is the so-called \emph{spherical equivalency}~\cite{2012_PUCCI1,2013_PUCCI,pucciPhd,phms15}, 
and 
it can be applied to NACA airfoils. Once applied,
control laws stabilizing reference velocities can be designed. These laws extend the vectored--thrust control paradigm \cite{ghm05,hhms09,naldi2017robust}  developed for systems with orientation-independent external forces
to 
orientation-dependent external forces.

\newpage
The paper is organized as follows. Section~\ref{sec:background} provides the notation, the background, and some definitions.
In Section~\ref{sec:existenceEquilibrium}, we address the problem of the existence of the equilibrium orientation, and in 
Section~\ref{sec:multiplicity} we study the multiplicity of this orientation in the case of NACA airfoils. The spherical equivalency and 
its relation to the  uniqueness of an equilibrium orientation is presented in Section~\ref{sec:equivalency}. In this section, we also present the application of the
spherical equivalency to NACA airfoils.
The spherical equivalency  is used in Section~\ref{sec:control} to propose a feedback control design method applicable to several vehicles. Simulation results for the airfoil NACA 0021 performing a transition maneuver between low velocity hovering and high velocity cruising are reported in 
Section~\ref{sec:simulations}. Remarks and perspectives conclude the paper.

\section{Background}
\label{sec:background}

\subsection{Notation}
\label{sec:notation}
\begin{figure}[tb]
\centering
\vspace*{0.4cm}
 \def\svgwidth{0.8\linewidth}
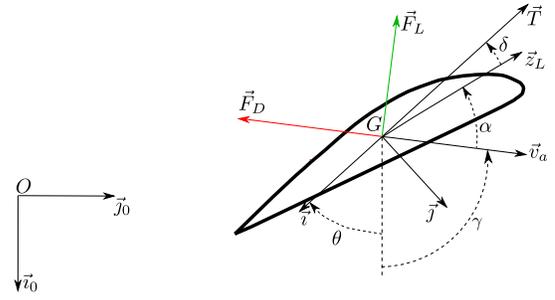
\caption{Propelled vehicle subject to aerodynamic forces.}
\label{fig:aerodyEff}
\end{figure}

\begin{itemize}
\item  The $i_{th}$ component of a vector $x\in \mathbb{R}^n$ is denoted as $x_i$.
\item $\mathcal{I} = \{O;\vec{\imath}_0,\vec{\jmath}_0\}$ is a fixed inertial frame with respect to (w.r.t.) which the vehicle's absolute pose is measured.
\item $\mathcal{B} = \{G;\vec{\imath},\vec{\jmath}\}$ is a frame attached to the body, and~$\vec{\imath}$ is parallel to the thrust force $\vec{T}$. This leaves two possible and opposite directions for $\vec{\imath}$. The direction chosen here
is consistent with the convention used for VTOL vehicles.
\item For the sake of brevity, $(x_1 \vec{\imath} + x_2 \vec{\jmath})$ is written as $(\vec{\imath},\vec{\jmath})x$.
\item $\{e_1,e_2\}$ is the canonical basis in $\mathbb{R}^2$, and $I$ is the $(2 \times 2)$ identity matrix.
\item $\vec{x} \cdot \vec{y}$ denotes the scalar product of two vectors $\vec{x}, \vec{y}$. 
\item  Given a function of time $f: \mathbb{R} {\rightarrow} \mathbb{R}^n$,  its first time derivative is denoted
as $\frac{d}{dt}f = \dot{f}$. Given a function $f$ of several variables, its partial derivative
w.r.t. some of them, say $x$, is denoted as $ \partial_x f {=} \frac{\partial f}{\partial x}$. 
Given a function $f(x):\mathbb{R}\rightarrow\mathbb{R}$, the first and second order derivative w.r.t. $x$ can be denoted by $f'$ and $f''$, respectively.
\item $G$ is the body's center of mass and 
\emph{m} is the (constant) mass of the vehicle.
\item $\vec{p}:= \vec{OG} = (\vec{\imath}_0,\vec{\jmath}_0)x$ denotes the body's position.
$\vec{v} =\frac{d}{dt}\vec{p}= (\vec{\imath}_0,\vec{\jmath}_0)\dot{x}= (\vec{\imath},\vec{\jmath})v$ denotes the
the body's linear velocity, 
and $\vec{a}=\frac{d}{dt}\vec{v}$ the linear acceleration.
\item The vehicle orientation is given by the angle $\theta$  between $\vec{\imath}_0$ and $ \vec{\imath}$. 
The rotation matrix of  $\theta$ is $R(\theta)$. The column vectors of $R$ are the vectors of coordinates of
$\vec{\imath},\vec{\jmath}$  in $\mathcal{I}$. The matrix $S = R(\pi/2)$ is a unitary skew-symmetric matrix. 
The body's angular velocity is $\omega := \dot{\theta}$.

\end{itemize}

\subsection{Equations of motion}
\label{sec:systemmodeling}

We consider two control inputs to derive the equations of motion: a {\em thrust} force $T$ 
along the body fixed direction~$\vec{\imath}$ ($\vec{T} = -T\vec{\imath}$), whose main role is to produce longitudinal motions, and 
a torque actuation, typically created via secondary propellers, rudders or flaps, etc. 
We assume that any desired torque can be produced so that the vehicle's angular velocity $\omega$ 
can be
used as a control variable.
 In the language of Automatic Control, 
this is a  \emph{backstepping} assumption, and producing the  angular velocity 
can be achieved via classical nonlinear techniques~\cite[p. 589]{kh02}. In the language of Aircraft Flight Dynamics, instead, 
this is the assumption of the \emph{guidance loop}, which focuses on the problem of determining the thrust intensity and the 
vehicle's orientation  to track a  desired reference position/velocity. 

The external forces acting on the body are assumed to be composed of the gravity $mg\vec{\imath}_0$ and the aerodynamic
forces $\vec{F}_a$. Applying Newton's law yields:
\begin{IEEEeqnarray}{RCL}
	 \label{eq:dynamics}
	 \IEEEyesnumber
	 m\vec{a} &=& mg\vec{\imath}_0+ \vec{F}_a - T\vec{\imath}, 
 	\IEEEyessubnumber 
 	\label{eq:newton2D} \\
	\dot{\theta} &=& \omega,
	\label{eq:dyncooM} \IEEEyessubnumber
\end{IEEEeqnarray}
with $g \in \mathbb{R} $ the gravitational acceleration. 
\subsection{Aerodynamic forces}
\label{par:aerodyForce}
\emph{Steady} aerodynamic forces at constant 
Reynolds and Mach numbers
can be written as follows~\cite[p. 34]{2010_AND}
\begin{IEEEeqnarray}{RLL}
	\label{aerodynamicForce2D}
	\vec{F}_a &=& k_a|\vec{v}_a|\left[c_L(\alpha)\vec{v}^\perp_a -c_D(\alpha)\vec{v}_a\right],
\end{IEEEeqnarray}
with $k_a {:=}\tfrac{\rho \Sigma}{2} $, $\rho$ the \emph{free stream} air density, 
$\Sigma$ the characteristic surface of the vehicle's body, 
$c_L(\cdot)$ the \emph{lift coefficient}, $c_D(\cdot)>0$  the \emph{drag coefficient} ($c_L$ and $c_D$ are called
\emph{aerodynamic characteristics}), $\vec{v}_a = \vec{v}-\vec{v}_w$ the \emph{air velocity},    $\vec{v}_w$  the wind's velocity,
$\vec{v}^\perp_a$ obtained by rotating the vector $\vec{v}_a$ by $90^\circ$ anticlockwise, i.e.  
	$\vec{v}^\perp_a = v_{a_1}\vec{\jmath} - v_{a_2}\vec{\imath}$,
and $\alpha$ the angle of attack. This latter variable is  
here 
defined as the angle between the
body-fixed {\em zero-lift} direction~$\vec{z}_L$, along which the airspeed does not produce lift forces, 
and the airspeed vector $\vec{v}_a$, i.e. 
\begin{equation}
	\alpha := \text{angle}(\vec{v}_a,\vec{z}_L).
	\label{eq:angleOfAttacVec}
\end{equation}
The model~\eqref{aerodynamicForce2D} neglects the so-called \emph{unsteady aerodynamics}, e.g., the flow pattern effects induced by \emph{fast} angular velocity motions~\cite[p.199]{2004_STENGEL}. This assumption is commonly accepted in the  literature dealing with large flight-envelope aircraft control~\cite{naldi2011optimal,casau2013hybrid,7039480,jung2012development,4803783,ccetinsoy2012design,muraoka2009quad,
flores2013transition,muraoka2009quad,frank2007hover,maqsood2010optimization,itasse2011equilibrium}: in fact, building global, unsteady aerodynamic models employable for control design purposes is still a  challenge for the specialized aerodynamic literature~\cite{bruntonPhd,brunton2014state}.

Now, denote the constant angle between the zero-lift direction $\vec{z}_L$ and the thrust $\vec{T}$  as $\delta$, i.e. 
$\delta := \mathrm{angle}(\vec{z}_L,\vec{T}),$
and  the angle between 
the gravity $g\vec{\imath}_0$ and~$\vec{v}_a$ as $\gamma$, i.e. 
$	\gamma := \mathrm{angle}(\vec{\imath}_0,\vec{v}_a).$ Then
(see Figure~\ref{fig:aerodyEff}):
\begin{IEEEeqnarray}{RCL}
	\alpha  &=& \theta -\gamma  +(\pi -\delta)
	\label{eq:angleOfAttack}, \\
	\IEEEyesnumber
	\label{eq:VaComponents}
	v_{a_1} &=& -|\vec{v}_a|\cos(\alpha+\delta)  \quad \ \ 
	\label{eq:va1}  \IEEEyessubnumber  \\
	v_{a_2} &{=}&  \hspace{0.34cm} |\vec{v}_a|\sin(\alpha+\delta). \IEEEyessubnumber
	 \label{eq:va2}
\end{IEEEeqnarray}

\subsubsection{Symmetric shapes}
\label{sec:shape}
\begin{figure}[t]
  \centering
 \def\svgwidth{0.71\linewidth}
        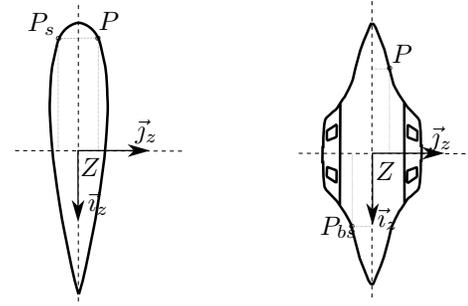
   \caption{Examples of symmetric and bisymmetric bodies.}
   \label{fig:symAndBisym}
   \vspace{-0.4cm}
\end{figure}
To characterize two kinds of  shape symmetries and their properties, let $\mathcal{B}_z~=~\{Z;\vec{\imath}_z,\vec{\jmath}_z\}$ be an orthonormal frame, and 
$P$ a point of the body surface~$\mathcal{S}$ -- see Figure~\ref{fig:symAndBisym}. 
Consider the vector~$\vec{ZP}$ and its expression w.r.t. the frame $\mathcal{B}_z$, i.e.
$\vec{ZP} :=  x\vec{\imath}_z+y\vec{\jmath}_z, \text{ with }  x,y \in~\mathbb{R}.$
Then, \emph{symmetric} and \emph{bisymmetric} shapes 
satisfy what follows.
\begin{hypothesis}[\textbf{Symmetry}]
\label{hy:symmetries2D} 
There exists a choice for the frame $\mathcal{B}_z$ such that the point $P_s$ defined by the vector 
$\vec{ZP}_s = x\vec{\imath}_z-y\vec{\jmath}_z$ 
belongs to $\mathcal{S}$ for any point $P$ of the  surface $\mathcal{S}$. Then, the  shape is said to be \emph{symmetric},
with axis of symmetry given by $\{Z,\vec{\imath}_z\}$.
\end{hypothesis}
\begin{hypothesis}[\textbf{Bisymmetry}]
\label{hy:bisymmetries2D}
There exists a choice for the frame $\mathcal{B}_z$ such that the point $P_{bs}$ defined by 
$\vec{ZP}_{bs} = -x\vec{\imath}_z-y\vec{\jmath}_z$ 
belongs to $\mathcal{S}$ for any point $P$ of the  surface $\mathcal{S}$. Then, the  shape is said to be \emph{bisymmetric},
with axes of symmetry given by $\{Z,\vec{\imath}_z\}$ and $\{Z,\vec{\jmath}_z\}$.
\end{hypothesis}
We assume that an axis of symmetry identifies two zero-lift-directions. Then, 
we choose the zero-lift-direction $\vec{z}_L$ in~\eqref{eq:angleOfAttacVec}  parallel to an axis of symmetry, which implies that 
$c_L(0)=c_L(\pi)=0$. Note that this choice still leaves two possible and opposite directions for 
the definition of the vector $\vec{z}_L$, which in turn may reflect
in two possible values
of the angle~$\delta$. 
Without loss of generality, the direction here chosen is that minimizing the angle~$\delta$. 

In light of the above  choice,
a symmetric shape induces aerodynamic
 characteristics $c_D( \alpha)$ and $c_L( \alpha)$ that are even and odd functions, respectively. 
\begin{property}
\label{propertySymmetricShape}
If the body shape $\mathcal{S}$ is symmetric and the zero-lift-direction $\vec{z}_L$ 
is parallel to the axis of symmetry, then 
hold:
\begin{IEEEeqnarray}{RCL}
	\IEEEyesnumber
	    \label{LiftPropertiesSymmetricB}
 		c_D( \alpha) &=& c_D( -\alpha),   \quad \quad c_L( \alpha) = -c_L( -\alpha), \quad \forall \alpha, \IEEEyessubnumber \IEEEeqnarraynumspace \\ 
		c_L( 0) &=& c_L( \pi) = 0. \IEEEyessubnumber \label{cLZeroCLPiUZ} 
		\IEEEeqnarraynumspace
\end{IEEEeqnarray}
\end{property}
Bisymmetric shapes have an additional symmetry about the axis $\vec \jmath_z$, thus implying the invariance
of the aerodynamic forces w.r.t. body rotations of $\pm \pi$. Then, 
the aerodynamic characteristics of bisymmetric shapes are $\pi-$periodic functions versus the angle 
$\alpha$. 
\begin{property}
\label{propertyBisymmetricShape}
If the body shape $\mathcal{S}$ is bisymmetric and the zero-lift-direction $\vec{z}_L$ 
is parallel to an axis of symmetry, then the aerodynamic coefficients satisfy~\eqref{LiftPropertiesSymmetricB} and
\begin{IEEEeqnarray}{RCL}
	    \label{LiftPropertiesBSymmetricB}
		c_D( \alpha) &=& c_D( \alpha \pm \pi), \quad \quad
		c_L( \alpha) = c_L( \alpha \pm \pi), \quad \forall \alpha.  \IEEEeqnarraynumspace 
\end{IEEEeqnarray}
\end{property}

\subsection{Problem statement and preliminary definitions}
The control objective is the asymptotic stabilization of  a reference velocity.
Let $\vec{v}_r(t)$ denote the differentiable reference velocity, and  $\vec{a}_r(t)$ its  time derivative, 
i.e. $\vec{a}_r(t)~=~\dot{\vec{v}}_r(t)$. Now,
define the velocity error as follows
\begin{IEEEeqnarray}{rcl}
  \label{velocityError}
  \vec{e}_v := \vec{v}-\vec{v}_r.
\end{IEEEeqnarray} 
Using System~\eqref{eq:dynamics} one obtains the following error model
\begin{IEEEeqnarray}{rCL}
	\label{eq:errorsDynamics}
	m\dot{\vec{e}}_v &=& \vec{F}-T\vec{\imath},
	\quad \quad \quad
	\dot{\theta} = \omega,
\end{IEEEeqnarray}
with $\vec{F}$  the \emph{apparent external force} defined by
\begin{IEEEeqnarray}{RCL}
  \label{externalApparentForce}
  \vec{F} &:=& m\vec{g} +\vec{F}_a-m\vec{a}_r. 
\end{IEEEeqnarray}
 Eq.~\eqref{eq:errorsDynamics} indicates that the equilibrium condition $\vec{e}_v {\equiv} 0$ requires
   $T\vec{\imath}(\theta) = \vec{F}({\vec{v}_r(t),\theta,t}),  \forall t, $
which 
in turn 
implies  
\begin{IEEEeqnarray}{RCL}
\IEEEyesnumber
  \label{equilibiumConditions}
  T &=& \vec{F}({\vec{v}_r(t),\theta,t})\cdot \vec{\imath}(\theta), \IEEEyessubnumber \label{equilibiumThrustCondition} \\
  0 &=& \vec{F}({\vec{v}_r(t),\theta,t}) \cdot \vec{\jmath}(\theta) \quad \forall t.  \IEEEyessubnumber \label{equilibiumOrinetationCo}
\end{IEEEeqnarray}
The existence of an orientation $\theta$ such that Eq.~\eqref{equilibiumOrinetationCo} is satisfied cannot be ensured \emph{a priori}. In fact,
the apparent external force $\vec{F}$ depends on the vehicle's orientation, and any change of this orientation affects both vectors 
$\vec{F}$ and $\vec{\jmath}$.
The dependence of $\vec{F}$ on the  orientation~$\theta$ comes from the dependence of the aerodynamic force $\vec{F}_a$ 
upon $\alpha$ (see Eqs.~\eqref{aerodynamicForce2D} and~\eqref{eq:angleOfAttack}). 
In view of Eq.~\eqref{equilibiumOrinetationCo}, we  state the definition next.
\begin{definition}
 \label{def:eqOrie}
 An \emph{equilibrium orientation} $\theta_e(t)$ 
 is a time function such that Eq.~\eqref{equilibiumOrinetationCo} is satisfied with $\theta =\theta_e(t)$. 
\end{definition}

\noindent
The \emph{existence} of an equilibrium orientation is a necessary condition for the asymptotic stabilization of a reference velocity.
In general, a reference velocity $\vec{v}_r(t)$ may induce several equilibrium orientations. 
To classify the \emph{number} of these equilibrium orientations, define the set $\Theta_{\vec{v}_r}(t)$ as
\begin{IEEEeqnarray}{RCL}
  \label{thetaSet}
 \Theta_{\vec{v}_r}(t) {:=} \bigg\{ \theta_e(t) {\in} \mathbb{S}^1 : 
 \vec{F}({\vec{v}_r(t),\theta_e(t),t}) \cdot \vec{\jmath}(\theta_e(t)) {=} 0 \bigg\}. \IEEEeqnarraynumspace
\end{IEEEeqnarray}
We further introduce a terminology 
when there exist only two opposite equilibrium orientations over large domains 
of 
the reference velocity 
$\vec{v}_r$.
\begin{definition}
 \label{def:genericallyUniqueCouple}
  We say that System~\eqref{eq:errorsDynamics} possesses a \emph{generically-unique equilibrium orientation} 
  if and only if there exists $\theta_e(t)$ such that 
 \begin{IEEEeqnarray}{RCL}
 \Theta_{\vec{v}_r}(t) = \bigg\{ \theta_e(t),\ \ \theta_e(t) + \pi  \bigg\}, \quad \forall t, \nonumber
 \end{IEEEeqnarray}
for any reference velocity $\vec{v}_r(t)$ except for a unique, continuous velocity $\vec{v}_b(t)$ such that 
 $\Theta_{\vec{v}_b}(t) = \mathbb{S}^1 \quad \forall t.$
\end{definition}

\noindent
For the systems possessing a generically-unique equilibrium orientation, 
the reference velocity  \emph{can be tracked} with only two, opposite vehicle orientations at any time~$t$. 
This holds 
for any reference velocity except for a unique \emph{bad} reference velocity 
$\vec{v}_b$. At this bad reference velocity, any orientation is of equilibrium, i.e.~\eqref{equilibiumOrinetationCo} 
is satisfied for any  orientation $\theta$. 


Remark that given an equilibrium orientation $\theta_e(t)$, the thrust intensity $T$ at the equilibrium configuration is 
given by Eq.~\eqref{equilibiumThrustCondition}  with $\theta = \theta_e(t)$. 
The existence of an equilibrium orientation ensuring a positive thrust is
of particular importance, since positive-thrust limitations represent a common constraint when considering aerial vehicles.
To characterize this existence, define
\begin{IEEEeqnarray}{RCL}
 \label{positiveThrustThe}
 \Theta^+_{\vec{v}_r}(t) {:=} \bigg\{ &\theta_e(t)& \in \Theta_{\vec{v}_r}(t) : 
 \vec{F}({\vec{v}_r,\theta_e,t})\cdot \vec{\imath}(\theta_e) \geq 0 \bigg\}.
  \IEEEeqnarraynumspace
\end{IEEEeqnarray}

\section{Existence of the equilibrium orientation}
\label{sec:existenceEquilibrium}
We know from experience that airplanes do fly, so 
the equilibrium orientation must exist in most cases. 
One may conjecture that the existence of an equilibrium orientation follows from
aerodynamic properties that hold independently of the body's shape, 
alike the \emph{passivity}  of aerodynamic forces.
The next lemma, however, points out that the  aerodynamic force passivity is not sufficient
to assert the existence of an equilibrium orientation.
\begin{lemma}
    \label{lemma:dissipativityNotSufficient}
    The passivity of the aerodynamic force, i.e.
    \begin{IEEEeqnarray}{r}
	\vec{v}_a \cdot \vec{F}_a(\vec{v}_a,\alpha) \leq 0, \quad \forall (\vec{v}_a,\alpha), \IEEEeqnarraynumspace \label{dissipativity}
    \end{IEEEeqnarray}
    is not a sufficient condition for the existence of an equilibrium orientation.
\end{lemma}
The proof is given in the Appendix.
Another route that we may follow to conclude about the existence of an equilibrium orientation is by 
considering specific classes 
of body's shapes. 
\begin{theorem} 
	\label{th:existence}
	Assume that the  aerodynamic coefficients $c_L(\alpha)$ and $c_D(\alpha)$ are continuous 
	functions, and that the reference velocity is differentiable, 
	i.e. $\vec{v}_r(t) \in \mathbf{\bar{C}}^1$. 
\begin{enumerate}
		\item[i)] If the body shape is symmetric 
		and the thrust is parallel to the its axis of symmetry, 
		then there exist at least two equilibrium orientations 
		for any reference velocity, i.e.
		 \[ \emph{\text{cardinality}}(\Theta_{\vec{v}_r}(t)) \geq 2 \quad \forall t, \quad \forall \vec{v}_r(t) \in \mathbf{\bar{C}}^1. \]
		\item[ii)] If the body's shape is bisymmetric,
		then there exists at least one equilibrium orientation  ensuring a 
		positive-semidefinite thrust  
		for any 
		reference velocity, i.e.
		\[\emph{\text{cardinality}}(\Theta^+_{\vec{v}_r}(t)) \geq 1 \quad \forall t, 
		\quad \forall \vec{v}_r(t) \in \mathbf{\bar{C}}^1, \]
		whatever the (constant) angle $\delta$.
	\end{enumerate}
\end{theorem}

The proof is given in the Appendix. 
Item~$i)$ asserts that for symmetric body's shapes powered by a thrust force parallel to 
their axis of symmetry, e.g. $\delta=0$,
the existence of (at least) two equilibrium orientations is guaranteed for any reference velocity. 
Item~$ii)$ 
states that the bisymmetry of the  shape implies the existence of an 
equilibrium orientation independently of the thrust direction with respect to 
the body,
i.e. the angle $\delta$. Of most importance, this item points out that 
the shape's bisymmetry implies the existence of an equilibrium orientation inducing  a positive-semidefinite thrust intensity independently of reference trajectories. 

Now, assume that the body's shape is symmetric and not bisymmetric. 
If the thrust force is not parallel to the shape's axis of symmetry, the assumptions of
Theorem~\ref{th:existence} are not satisfied and the existence of an equilibrium orientation cannot be asserted.
Yet,  common sense makes us think that an equilibrium orientation still exists.
By considering symmetric shapes, the next theorem states conditions ensuring the existence of an equilibrium orientation 
independently of reference velocities and thrust directions w.r.t. the body's zero-lift direction.

\begin{figure*}[t!]
  \centering
  \vspace*{-0.2cm}
  \def\svgwidth{0.8\linewidth}\input{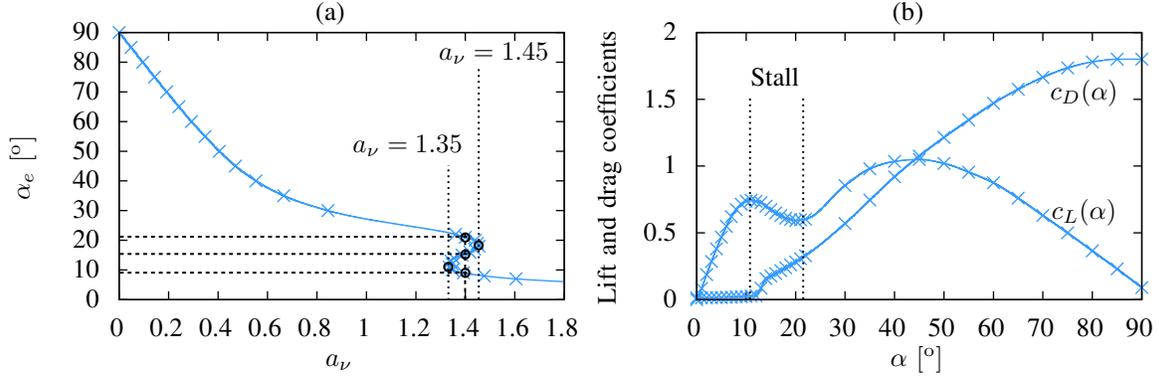}
  \caption{(a): pattern of the equilibrium angles $\alpha_e$; (b):  aerodynamic coefficients of NACA 0021 with a length $l = 0.91 m$, 
  a chord length $c = 0.1524m$, and 
Reynolds and Mach numbers equal to $R_e = 160 \cdot 10^3$ and $M = 0.3$, respectively.}
  \label{fig:bifurcationsAndStall}
\end{figure*}

\begin{theorem}
	\label{th:existenceDeltaDifZeroSymm}
	Consider symmetric shapes. Assume that the aerodynamic coefficients $c_L(\alpha)$ and $c_D(\alpha)$ 
	are continuous functions, and that $c_D(\pi) > c_D(0)$. 
	If there exists an angle $\alpha_s \in (0,\pi/2)$ such that $c_L(\alpha_s) > 0$ and 
	  \begin{IEEEeqnarray}{RCL}
		  \label{conditionMuNe0}
		  \tan(\alpha_s) &\leq& \frac{c_D(\alpha_s)-c_D(\pi)}{c_L(\alpha_s)},
	  \end{IEEEeqnarray}
	  then there exists at least one equilibrium orientation 
	  for any 
	  reference velocity, i.e.
		\[ \emph{\text{cardinality}}(\Theta_{\vec{v}_r}(t)) \geq 1 \quad \forall t, 
		\quad \forall \vec{v}_r(t) \in \mathbf{\bar{C}}^1, \]
		whatever the (constant) angle $\delta$ between the zero-lift direction and the thrust force.
\end{theorem}
The proof is given in the Appendix. 
The key hypothesis in Theorem~\ref{th:existenceDeltaDifZeroSymm} is the existence of an angle $\alpha_s$ such that the
condition~\eqref{conditionMuNe0} is satisfied. 
Seeking for this angle requires some aerodynamic data, and it may be airfoil and flow regime specific.
Recall, however, that  \emph{stall phenomena} 
(see, e.g., Figure~\ref{fig:bifurcationsAndStall}b) involve rapid, usually important, lift decreases and drag increases.
Then the likelihood of satisfying the condition~\eqref{conditionMuNe0} 
with $\alpha_s$ belonging to the stall region is very high. In fact, we verified that Theorem~\ref{th:existenceDeltaDifZeroSymm} applies 
with $\alpha_s$ belonging to the stall region for 
the NACA airfoils 0012, 0015, 0018, and 0021 at $M = 0.3$ and several Reynolds numbers (data taken from~\cite{CYBERIAD}).

In light of the above, an equilibrium orientation exists in most cases if the airfoil is 
quasi-symmetric, and this existence
is independent from the thrust direction relative to the body. From a control perspective,
however,
one is also interested in  the number of possible vehicle equilibrium orientations along the reference velocity.

\section{A case study: multiplicity of the equilibrium orientation in the case of NACA airfoils}
\label{sec:multiplicity}
This section studies the multiplicity of the equilibrium orientation and its control
consequences by considering the experimental aerodynamic 
coefficients of the symmetric airfoil NACA~0021 in steady, horizontal flight.
Recall that this multiplicity equals the cardinality of the set $\Theta_{\vec{v}_r}(t)$ 
given by~\eqref{thetaSet}.
For the sake of simplicity, we assume no wind, i.e.
	 $|\vec{v}_w| \equiv 0$,
a thrust force aligned with the zero-lift direction, i.e.
	 $\delta = 0$,
and a desired steady-horizontal flight, i.e. 
	$\dot{x}_r = \nu e_2$,
where $\dot{x}_r$ is the vector of coordinates of $\vec{v}_r$ expressed in the inertial frame.
For an analysis of the equilibrium orientation multiplicity along other flight directions see~\cite[p. 76]{pucciPhd}.

Since the airfoil NACA~0021 is symmetric, then
Theorem~\ref{th:existence} with $\delta = 0$ ensures the existence of at least one equilibrium  orientation along any reference velocity.  
Now, from the definition of the set $\Theta_{\vec{v}_r}$ given by~\eqref{thetaSet}, the cardinality of this set equals the number of 
solutions~$\theta_e$ to the following equation
\begin{IEEEeqnarray}{RCL}
   F(\dot{x}_r,\theta_e)^T R(\theta_e)e_2 = 0, 
   \label{eq:wnugeneralPre}
\end{IEEEeqnarray}
where 
$F$ is the vector of coordinates of $\vec{F}$, given by~\eqref{externalApparentForce}, expressed in the inertial frame, 
i.e. 
 \[ F = mge_1 +k_a|\dot{x}_r|\left[ c_L(\alpha_e)S-c_D(\alpha_e)I\right]\dot{x}_r,\]
with 
\[\alpha_e = \theta_e -\gamma_r +\pi,\] and $\gamma_r = \text{angle}(e_1,\dot{x}_r) = \pi/2$.
By replacing 
	$\theta_e = \alpha_e +\gamma_r-\pi$
in Eq.~\eqref{eq:wnugeneralPre}, one gets
\begin{IEEEeqnarray}{RCL}
	 \left[1{-}a_{\nu}c_L(\alpha_e)\right]\cos(\alpha_e){-}a_{\nu}c_D(\alpha_e)\sin(\alpha_e) &=& 0, \IEEEeqnarraynumspace
	 \label{eq:wCruiseNoWind}
\end{IEEEeqnarray}
where $a_\nu$ is a \emph{dimensionless number} defined by 
	\[a_\nu~:=~\frac{k_a\nu^2}{mg}.\]
 Therefore, the problem of seeking for the equilibrium orientations $\theta_e$ 
is \emph{equivalent} to the problem of finding the \emph{equilibrium angles of attack} 
$\alpha_e$ that satisfy~\eqref{eq:wCruiseNoWind}. 
Observe that from~\eqref{eq:wCruiseNoWind},
we can find the explicit expression of the parameter $a_\nu$  in function of the equilibrium 
angles 
$\alpha_e$, 
\begin{IEEEeqnarray}{RCL}
	 a_\nu(\alpha_e) &=& \frac{\cot(\alpha_e)}{c_D(\alpha_e)+c_L(\alpha_e)\cot(\alpha_e)}. \IEEEeqnarraynumspace
	 \label{eq:a(ae)}
\end{IEEEeqnarray}
A picture of the equilibrium angle $\alpha_e$ as a 
function of the cruise velocity $\nu$ is then obtained by plotting Eq.~\eqref{eq:a(ae)} along the angle of attack $\alpha_e$. 
Figure~\ref{fig:bifurcationsAndStall}a 
depicts the function~\eqref{eq:a(ae)} evaluated with the experimental aerodynamic 
characteristics shown in Figure~\ref{fig:bifurcationsAndStall}b. 
From this figure we see that 
the equilibrium angle of attack is unique 
as long as the parameter $a_\nu < 1.35$. At $a_\nu = 1.35$, 
the equilibrium angle of attack bifurcates in multiple points. The local bifurcation of $\alpha_e$ is a \emph{saddle-node} kind 
since a couple of equilibrium angles \emph{collide and annihilate each other}  \cite{1996_CHOW,1990_WIGGINS}  when crossing the bifurcation values $a_\nu = 1.35$ and $a_\nu = 1.45$. 
When $a_\nu$ belongs to a neighborhood of $1.4$, three equilibrium angles of attack arise, and a steady-horizontal 
flight may  theoretically be performed with three different vehicle's orientations.

The bifurcation analysis of the equilibrium orientations is beyond the scope of the present paper, all the more so because 
these local phenomena occur principally on the highly nonlinear and chaotic stall region.
Let us just remark that stall phenomena -- intended as a lift decrease when $\alpha \in (0,45^\circ)$ -- do not
always imply a bifurcation of the equilibrium orientation~\cite[Lemma 7.5, p. 74]{pucciPhd}.

\subsection{Ill-conditioning of the control problem for constant height transition maneuvers}
\label{subsec:illcond}
\begin{figure}[t]
  \centering
  \vspace*{-0.4cm}
   \def\svgwidth{0.8\linewidth}\input{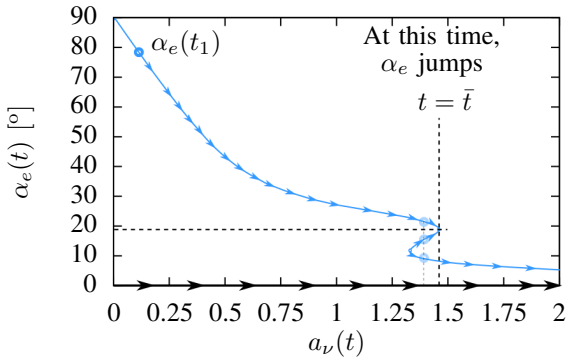}
  \caption{Equilibria pattern variation for  $\dot{x}_r = (0,\nu t)^T$.}
  \label{fig:bifurcationsDin}
\end{figure} 
A consequence of the existence of several equilibria is that given a continuous reference velocity, 
the associated equilibrium orientation $\theta_e(t)$ may be discontinuous. In this case, 
the reference velocity cannot be \emph{perfectly tracked}.
Also, the continuity 
of the equilibrium orientation $\theta_e(t)$ is a necessary condition for the  
asymptotic stabilization of  the equilibrium
$\vec{e}_v = 0 $ associated with System~\eqref{eq:errorsDynamics}. In fact, 
the control input $\omega$ at the equilibrium, i.e. $\omega = \dot \theta_e(t)$, must be defined for any~$t$, and this is not the case
when $\theta_e(t)$ is discontinuous.

The fact that the continuity of the reference velocity does not in general imply the continuity of the equilibrium 
orientation $\theta_e(t)$ 
is visually shown in Figure~\ref{fig:bifurcationsDin}. This figure depicts the time evolution of the
equilibrium angle of attack when considering constant-height {\em transition maneuvers} between
hovering and high-velocity cruising, 
\begin{equation}
	\dot{x}_r(t) = \nu t(0,1)^T,
	\label{eq:cruisingFlightConstantAcc}
\end{equation}
with $\nu$ a (small) positive number.  On the time interval $t \in (0,t_1)$ (see Figure~\ref{fig:bifurcationsDin}), one has $\alpha_e \approx 90^\circ$ because the horizontal reference velocity is of low intensity (the thrust opposes the weight). As time goes by, the intensity of the reference velocity increases, and this in turn implies smaller values of the 
angle of attack at the equilibrium configuration. At $t = \bar{t}$, the equilibrium attitude $\alpha_e(t)$ instantaneously goes from 
$19^\circ$ to $8^\circ$, thus making the equilibrium orientations $\theta_e(t)$ discontinuous. 
Such discontinuities destroy the \emph{well-posedness} of the asymptotic stabilization problem related to the transition maneuver 
given by \eqref{eq:cruisingFlightConstantAcc}, and may cause a LOC if not carefully dealt with. Then, the reference velocity~\eqref{eq:cruisingFlightConstantAcc}
cannot be perfectly tracked by any aircraft whose aerodynamic characteristics are similar to those shown in 
Figure~\ref{fig:bifurcationsAndStall}b. This calls for specific studies on the planning of desired reference trajectories representing 
transition maneuvers between hovering and high-velocity cruising flight,
which are, however,
beyond the scope of the present paper.

\section{Uniqueness of the equilibrium orientation and spherical equivalency}
\label{sec:equivalency}

In the previous section, we have seen that a reference velocity $\vec{v}_r$ may be  perfectly tracked with several equilibrium 
orientations. This poses two interesting problems. 

\vspace{0.3cm}
\noindent
\emph{Global uniqueness: can we 
find aerodynamic models inducing a unique equilibrium orientation for any given reference velocity?} 

\vspace{0.3cm}
\noindent
\emph{Local uniqueness: 
can we find conditions ensuring that the equilibrium orientation is isolated for a given reference velocity and aerodynamic model?} 

\vspace{0.3cm}
We shall see  that the  uniqueness of the equilibrium orientation is related to the possibility of transforming  
system~\eqref{eq:errorsDynamics} -- either globally or locally -- into an equivalent meaningful system. This transformed system  
is of the same form of~\eqref{eq:errorsDynamics}, but the transformed apparent external force has a direction independent of the vehicle's orientation, as in the case of spherical shapes. 
The transformation is  called  \emph{spherical equivalency}.

\subsection{Global spherical equivalency and  generically-unique equilibrium orientation}
To introduce the spherical equivalency, define 
\begin{IEEEeqnarray}{RCL}
	\label{clBcdB} 
\IEEEyesnumber
	  \bar{c}_L(\alpha,\lambda) &:=&  c_L(\alpha) -\lambda\sin(\alpha+\delta), \IEEEyessubnumber \label{clB}  \\
	  \bar{c}_D(\alpha,\lambda) &:=&  c_D(\alpha) +\lambda\cos(\alpha+\delta),  \IEEEyessubnumber \label{cdB}
\end{IEEEeqnarray}
with $\lambda \in \mathbb{R}$. In view of $\vec{v}^\perp_a = v_{a_1}\vec{\jmath} - v_{a_2}\vec{\imath}$ and~\eqref{eq:VaComponents}, one can verify that $\vec{F}_a$ given by~\eqref{aerodynamicForce2D} 
can be decomposed as follows
\begin{IEEEeqnarray}{RCL}
\vec{F}_a &=& k_a |\vec{v}_a|\left[ \bar{c}_L \vec{v}^\perp_a-\bar{c}_D\vec{v}_a\right] -\lambda k_a |\vec{v}_a|^2\vec{\imath}.  
\nonumber
\label{eq:FaNewDecomposition} \IEEEeqnarraynumspace
\end{IEEEeqnarray}
Then, 
the dynamics of the velocity errors~\eqref{eq:errorsDynamics} become
\begin{IEEEeqnarray}{RCL}
	\label{eq:dynamicsVelocityErrorNF}
	m\dot{\vec{e}}_v &=& \vec{F}_p-T_p\vec{\imath},  
	\IEEEyessubnumber  \\
	\dot{\theta} &=& \omega,  
	\IEEEyessubnumber  
\end{IEEEeqnarray}
with 
\begin{IEEEeqnarray}{RCL}
  \label{FpTp}
  \IEEEyesnumber
    \vec{F}_p &:=&mg\vec{\imath}_0 + \vec{f}_p -m\vec{a}_r,
  \IEEEyessubnumber   \label{eq:FpExt} \IEEEeqnarraynumspace \\
    \vec{f}_p &:=& k_a |\vec{v}_a|\left[ \bar{c}_L(\alpha,\lambda) \vec{v}^\perp_a{-}\bar{c}_D(\alpha,\lambda)\vec{v}_a\right],
  \IEEEyessubnumber   \label{eq:fpExt} \IEEEeqnarraynumspace \\
T_p &:=& T + k_a \lambda|\vec{v}_a|^2. \IEEEyessubnumber \label{eq:Tp} 
\end{IEEEeqnarray}
In light of the above, one has the following result.	
\begin{lemma}
\label{lemma:conditionLS}
Assume that the aerodynamic coefficients are twice differentiable functions.
	\begin{enumerate}
	\item[i)] 
  System~\eqref{eq:errorsDynamics} can be transformed into the form~\eqref{eq:dynamicsVelocityErrorNF} 
  with $\vec{F}_p$ independent of $\theta$, $\forall  \vec{v}_a$, if and only if 
\vspace{0.2cm}
   \begin{IEEEeqnarray}{lCL}
\hspace{-1cm}
\vspace{0.2cm}
	(c_D'' -2c_L') \sin(\alpha+\delta) {+}(c_L''+2c'_D)\cos(\alpha+\delta) = 0. \IEEEeqnarraynumspace 
	\label{eq:condition}
  \end{IEEEeqnarray}
  In this case, the function
  $\lambda$ in~\eqref{clBcdB}-\eqref{FpTp}  is given by
	\begin{IEEEeqnarray}{rCL}
	\lambda(\alpha) = c'_L\cos(\alpha+\delta) + c'_D\sin(\alpha+\delta).
	\label{eq:pGC}
\end{IEEEeqnarray}
	\item[ii)] Assume that the thrust force is parallel to the zero-lift direction $\vec{z}_L$ so as $\delta = 0$. If 
\begin{IEEEeqnarray}{RCL}
	\bar{c}_D = c_D(\alpha) + c_L(\alpha) \cot(\alpha)
		\label{eq:condSimpler}
\end{IEEEeqnarray}
is a constant number, then	
	\begin{IEEEeqnarray}{rCL}
	\label{eq:pPC}
	\lambda(\alpha) = \frac{c_L(\alpha)}{\sin(\alpha)}
\end{IEEEeqnarray}
allows  to transform   System~\eqref{eq:errorsDynamics} into the form~\eqref{eq:dynamicsVelocityErrorNF} 
  with $\vec{F}_p$ independent of $\theta$.
	\end{enumerate}
\end{lemma}
The proof is given in the Appendix. 
Item i) states a necessary and sufficient condition on the aerodynamic coefficients that defines the cases when 
the vector $\vec{F}_p$, evaluated with $\lambda$ as~\eqref{eq:pGC}, 
is independent of the vehicle's orientation.
A simpler, only sufficient condition that allows  for the aforementioned transformation 
is stated in  item ii) when the thrust force is
parallel to the zero lift direction. More precisely, if the condition~\eqref{eq:condSimpler} is satisfied, 
one has $\bar{c}_L~=~0$ in~\eqref{clBcdB}. Then, the 
equivalent aerodynamic force $\vec{f}_p$ in~\eqref{eq:fpExt} is reduced to drag forces, i.e.
\[\vec{f}_p {=} -k_a |\vec{v}_a|\bar{c}_D\vec{v}_a.\]
This means that the shapes whose aerodynamic coefficients 
satisfy~\eqref{eq:condSimpler} with $\bar{c}_D > 0$ can be viewed as spheres once the variable change~\eqref{eq:Tp} 
in the thrust  is applied.

When the conditions in Lemma~\ref{lemma:conditionLS} are satisfied,
one can inherit the equilibria analysis and the control design developed for
spherical shapes but applied to the transformed system~\eqref{eq:dynamicsVelocityErrorNF}.
In fact, the following result holds.
\begin{lemma}
  \label{UnicityGeneric}
  Assume that the aerodynamic coefficients satisfy the condition~\eqref{eq:condition}, and that $\bar{c}_D$, given 
  by~\eqref{cdB}-\eqref{eq:pGC}, is a positive-constant.
  Then System~\eqref{eq:errorsDynamics}
  possesses a generically-unique equilibrium orientation given by 
\begin{IEEEeqnarray}{rCL}
	\theta_e(t) = \text{\emph{angle}} \left(\vec{\imath}_0,\vec{F}_p(\vec{v}_r(t),t)\right).
	\label{thetaeTran}
\end{IEEEeqnarray}
\end{lemma}
The proof is given in the Appendix. The above lemma highlights that the aerodynamic models
satisfying the condition~\eqref{eq:condition} induce an explicit expression for the equilibrium orientation $\theta_e$. 
Furthermore,
this equilibrium orientation is unique in the sense of definition~\ref{def:genericallyUniqueCouple}.

\subsection{Spherical equivalent shape of  NACA airfoils}
\label{sec:applicationNaca}
We show in this section that there exist aerodynamic models capable of approximating experimental data taken for NACA airfoils and satisfying the conditions~\eqref{eq:condition}-\eqref{eq:condSimpler}.
\newpage
\begin{proposition}
\label{prop:specificModels}
The following results hold.
	\begin{enumerate}
		\item[i)] The modeling functions 
	  \begin{equation}
		\label{coeffSmallReynolds1}
		\begin{system}
            c_L(\alpha) &=&  c_1\sin(2\alpha) \\
            c_D(\alpha) &=& c_{0} + 2c_1\sin^2(\alpha),
		\end{system}
	  \end{equation}
	  satisfy the condition \eqref{eq:condition} and yield
	  \begin{IEEEeqnarray}{RCLRLL}
	  		 \hspace*{-0.4cm}
			\bar{c}_L &=&  -c_1\sin(2\delta),
			\quad \quad \quad
			\lambda&= &2 c_1 \cos(\alpha - \delta)
		 \nonumber \IEEEeqnarraynumspace\\
		 \hspace*{-1cm}
		 \bar{c}_D &=& c_{0} + 2c_1\cos^2(\delta), 
		 \quad
		 T_p &=& T + 2c_1k_a|\vec{v}_a|^2\cos(\alpha{-}\delta).  
		 \IEEEeqnarraynumspace \label{TpIdeal1} \nonumber
	  \end{IEEEeqnarray}
\item[ii)]  If the thrust force $\vec{T}$ is parallel to the zero-lift direction $\vec{z}_L$ so as $\delta = 0$, then the modeling functions 
\begin{IEEEeqnarray}{rCl}
	\begin{system}
	c_{L}(\alpha) &=&  \frac{0.5c^2_2}{(c_2-c_3)\cos^2(\alpha)+c_3}\sin(2\alpha)  \label{eq:clSa}\\
	c_{D}(\alpha) &=& c_{0} + \frac{c_2c_3}{(c_2-c_3)\cos^2(\alpha)+c_3}\sin^2(\alpha), 
	\end{system} \IEEEeqnarraynumspace
	\label{coeffLargeReynoldsSamllAlpha1}
\end{IEEEeqnarray}
	  satisfy the condition \eqref{eq:condSimpler} and yield
	  \begin{IEEEeqnarray}{RCLLL}
			\bar{c}_L &=&  0, \quad \quad \quad \quad \lambda &= &\tfrac{c^2_2 \cos(\alpha)}{(c_2 - c_3)\cos^2(\alpha) + c_3} \nonumber \\
            \bar{c}_D &=&  c_{0} + c_2, \quad   T_p &=& T + \tfrac{c^2_2k_a|\vec{v}_a|^2\cos(\alpha)}{(c_2-c_3)\cos^2(\alpha)+c_3}.  \nonumber 
  \label{TpIdeal2} 
	  \end{IEEEeqnarray}
\end{enumerate}
\end{proposition}
\noindent
Concerning the functions~\eqref{coeffSmallReynolds1}, we show the following result.
\begin{lemma}
 \label{uniquenessBeautifulModel}
 Consider symmetric shapes. The model~\eqref{coeffSmallReynolds1} is the only family of aerodynamic coefficients independent of 
 $\delta$ that yield
  a vector $\vec{F}_p$ independent of $\theta$ 
  whatever
   the angle~$\delta$.
\end{lemma}
The proof is in the Appendix. The process of approximating experimental  data with the functions~\eqref{coeffSmallReynolds1} 
is shown in  Figure~\ref{fig:Approximation21NoStall}
(NACA 0021 with Mach and Reynolds numbers equal to $(R_e,M) \approx (160 \cdot 10^3, 0.3)$~\cite{CYBERIAD}).
 For this example, the identified coefficients are $c_0=0.0139$ and $c_1=0.9430$.
The approximation result, although not perfect, should be sufficient for control design purposes
at small Reynolds numbers~$R_e$
--~e.g. small-chord-length airfoils ~-- 
at
which stall phenomena are less pronounced~\cite{2010_Zhou}.  
In this respect, small vehicles are advantaged over large ones.
The model~\eqref{coeffSmallReynolds1}, in fact,  is reminiscent of the aerodynamic coefficients of a \emph{flat plate} when setting $c_0 = 0$~\cite{tangler2005}. When $\delta = 0$, we then speculate that a flat plate is equivalent to a sphere once the variable change
$T \rightarrow T_p$ 
in the thrust input is applied.

Figure~\ref{fig:Approximation21NoStall} shows that the experimental data 
 are basically independent of the Reynolds number when the angle of attack increases  beyond the \emph{stall region}~\cite{cory2008,moore2012}. Then, the higher the  Reynolds number, the worse  the  approximation result at small $\alpha$ only. 
In contrast, Figure~\ref{fig:smallAlpha} shows that the modeling functions~\eqref{coeffLargeReynoldsSamllAlpha1} yield better approximations at \emph{small} angles of attack independently of the Reynolds number. In fact, the second order Taylor expansions of the functions~\eqref{coeffLargeReynoldsSamllAlpha1} at $\alpha = 0$ is
$c_{L}(\alpha) =  c_2\alpha$, $c_{D}(\alpha) = c_{0} + c_3\alpha^2$,
which are the classical modeling functions used to approximate steady aerodynamic characteristics at low angles of attack~\cite{2004_STENGEL}. 
The quality of the  approximations 
provided by~\eqref{coeffLargeReynoldsSamllAlpha1}, however,
worsens when the angle of attack gets close to the \emph{stall region}.
\begin{figure}[p]
\vspace{-0.7cm}
 \centering
 \small{\input{figures/approxflatplate.tex}}
 \caption{Measurements and approximations of $c_L$ and $c_D$.}
 \label{fig:Approximation21NoStall}
 \small{\input{figures/smallalpha.tex}}
 \caption{Measurements and approximations  at \emph{small} $\alpha$.}
 \label{fig:smallAlpha}
 \small{\input{figures/approxcomplete.tex}}
\vspace{-0.3cm}
 \caption{Measurements and approximations of $c_L$ and $c_D$.}
 \label{fig:ApproximationComplete}
\end{figure}

Therefore,
we combine the models~\eqref{coeffLargeReynoldsSamllAlpha1} and~\eqref{coeffSmallReynolds1}  
to approximate the experimental data
 taken at large domains of $(R_e,\alpha)$. 

Consider, for instance, the following
smooth-rectangular function 
$\sigma(\cdot)$ defined by 
\begin{IEEEeqnarray}{rCl}
	\label{sigma}
	\sigma(\bar{\alpha},\bar{k},\alpha) = 
	\tfrac{1+\tanh(\bar{k}\bar{\alpha}^2-\bar{k}\alpha^2)}{1+\tanh(\bar{k}\bar{\alpha}^2)}, \quad \alpha \in [-\pi,\pi), \IEEEeqnarraynumspace
\end{IEEEeqnarray}
with $\bar{k},\bar{\alpha} \in \mathbb{R}$. 
This function is chosen so as to have $\sigma\approx1 $  at
small angles of attack, and $\sigma \approx 0$  at
large angles of attack. 
Let $(c_{L_L},c_{D_L})$ and $(c_{L_S},c_{D_S})$ denote the modeling functions given by 
Eqs.~\eqref{coeffSmallReynolds1} and~\eqref{coeffLargeReynoldsSamllAlpha1}, respectively. 
A combined model is then given by 
\begin{IEEEeqnarray}{rCl}
\IEEEyesnumber
	\label{CompleteModel}
	c_L(\alpha) &{=}&  c_{L_S}(\alpha)\sigma(\bar{\alpha},\bar{k}_L,\alpha) {+} c_{L_L}(\alpha)[1{-}\sigma(\bar{\alpha},\bar{k}_L,\alpha)]  \label{eq:clC} \IEEEyessubnumber \\
	c_D(\alpha) &{=}& c_{D_S}(\alpha)\sigma(\bar{\alpha},\bar{k}_D,\alpha) {+} c_{D_L}(\alpha)[1{-}\sigma(\bar{\alpha},\bar{k}_D,\alpha)].\IEEEyessubnumber 
	\IEEEeqnarraynumspace
\end{IEEEeqnarray}
Figure~\ref{fig:ApproximationComplete} depicts typical approximations given by the functions~\eqref{CompleteModel}. The estimated parameters at $R_e = 160 \cdot 10^3$ are
\begin{IEEEeqnarray}{rCl}
\label{coefficientsReL}
\begin{system}
	  (c_0,c_1,c_2,c_3) &=& (14\cdot 10^{-3},0.95,5.5,0.3) 
  \\
	  (\bar{\alpha },k_L,k_D) &=& (11^\circ, 28,167),
\end{system}
\end{IEEEeqnarray}
while at $R_e = 5 \cdot 10^6$ they are 
$c_0=0.0078$, 
$c_1=0.9430$, 
$c_2 = 6.3025 $, 
$c_3 = 0.1378$, 
$\bar{\alpha }= 18^\circ$, 
$k_L =   12$, and 
$k_D~=~86$. 
Figure~\ref{fig:ApproximationComplete} also shows that~\eqref{CompleteModel} can approximate  the main  aerodynamic coefficient variations including 
\emph{stall phenomena}.

\subsection{Local spherical equivalency and local uniqueness of the equilibrium orientation}
By construction of the model~\eqref{CompleteModel}, the force 
$\vec{F}_p$ in~\eqref{eq:FpExt}-\eqref{eq:pGC} is \emph{almost} independent of the vehicle's 
orientation if the angle of attack is away from the stall region, and $\delta = 0$. 
The control problem remains open when the reference velocity requires crossing this region and, more generally, 
when 
the aerodynamic model does not satisfy the condition~\eqref{eq:condition}.
This section shows that the transformed dynamics~\eqref{eq:dynamicsVelocityErrorNF}-~\eqref{eq:pGC} 
encompasses meaningful properties, which are instrumental for control control design purposes, \emph{independently}  of the aerodynamic forces. 
\begin{theorem}
\label{th:conditionLS}
  Assume that  there exists an equilibrium orientation~$\theta_e(t)$ for the reference velocity $\vec{v}_r(t)$, 
 and  that
  the aerodynamic coefficients are twice differentiable. Consider System~\eqref{eq:dynamicsVelocityErrorNF}-\eqref{FpTp} with $\lambda$ given by~\eqref{eq:pGC}. 
  If the vector $\vec{F}_p$ given by~\eqref{eq:FpExt} is different from zero at the  equilibrium point, i.e.
  \begin{IEEEeqnarray}{rCL}
	|\vec{F}_p(\vec{v}_r(t),\theta_e(t),t)| > 0,
	\label{eq:localUniqueness}
  \end{IEEEeqnarray}
 then  
 \begin{enumerate}
		\item[i)] the direction of  $\vec{F}_p$ is locally constant w.r.t. the vehicle's orientation at the equilibrium point, i.e.
	\begin{IEEEeqnarray}{rCL}
	  \partial_\theta \left[ \frac{\vec{F}_p}{|\vec{F}_p|} \right] \bigg|_{(\vec{e}_v,\theta)=(0,\theta_e(t))} =0;
	  \label{eq:magicFp}
	\end{IEEEeqnarray}
	\item[ii)]  the equilibrium orientation $\theta_e(t)$ is isolated and differentiable at the time $t$.
  \item[iii)] the linearization of~\eqref{eq:dynamicsVelocityErrorNF} at $(\vec{e}_v,\theta) {=} (0,\theta_e(t))$ 
  is controllable with $(T_p,\omega)$ taken as control inputs.
  
\end{enumerate}
\end{theorem}
The proof is in the Appendix.
The result~i) asserts that in a neighborhood of the equilibrium configuration, varying the 
thrust direction $\vec{\imath}$ does not perturb the 
direction of the vector~$\vec{F}_p$: this is why we say that system~\ref{eq:errorsDynamics} is locally
equivalent to a spherical shape system.
The condition $|\vec{F}_p| \ne 0$, in fact,  implies 
that the linearization of System~\eqref{eq:dynamicsVelocityErrorNF} at $(\vec{e}_v,\theta) = (0,\theta_e(t))$ is controllable (see the result~iii)) as in the cases where the external apparent force does not depend upon the vehicle orientation, e.g. spherical shape case.
\section{Control design}
\label{sec:control}

This section presents control laws for the asymptotic stabilization of the reference velocity $\vec{v}_r(t)$. 

Theorem~\ref{th:conditionLS}  points out that the direction of $\vec{F}_p$ in~\eqref{eq:dynamicsVelocityErrorNF}  is \emph{almost constant} close to the equilibrium configuration $(\vec{e}_v,\theta)=(0,\theta_e)$. 
A local control strategy then consists in aligning the thrust 
direction $\vec{\imath}$ with the direction of $\vec{F}_p$ 
and in opposing the magnitude $T_p$ to the intensity of $\vec{F}_p$.
\subsection{Equilibria of interest}
The transformed system~\eqref{eq:dynamicsVelocityErrorNF} shows that $\vec{e}_v \equiv 0$ implies
\begin{IEEEeqnarray}{RCLRCL}
  \label{equilibiumOrinetationConditionsLC2}
  T_p &=& \vec{F}_p({\vec{v}_r,\theta_e,t})\cdot \vec{\imath}(\theta_e), \quad 
  0 = \vec{F}_p({\vec{v}_r,\theta_e,t}) \cdot \vec{\jmath}(\theta_e), \quad \forall t.  \nonumber
\end{IEEEeqnarray}
Let $\tilde{\theta} \in (-\pi,\pi]$ denote the angle between $\vec{\imath}$ and $\vec{F}_p$. 
The control objective is then equivalent to the asymptotic stabilization of either 
$\tilde{\theta} {=} 0$ or  $\tilde{\theta} = \pi$, 
depending on the
equilibrium orientation $\theta_e$. These two equilibria correspond to either $T_p {=} |\vec{F}_p|$ or $T_p {=} {-}|\vec{F}_p|$, respectively. 
However, we derive control laws stabilizing $(T_p,\tilde{\theta}) = (|\vec{F}_p|,0)$ only. 
Let us   justify this choice. 
Assume a thrust parallel to the zero-lift direction so as $\delta {=} 0$. 
Eqs.~\eqref{eq:Tp}  and~\eqref{eq:pGC} point out that the  $T_p$ is given by
\[ T_p = T + k_a|\vec{v}_a|^2\left[ c'_L\cos(\alpha) + c'_D\sin(\alpha)\right].\]
Now assume also an equilibrium configuration at small velocities, i.e. the second term on the right hand side of the above equation is negligible. 
Then, stabilizing this reference velocity with a positive thrust $T$ requires a positive~ $T_p$, and consequently 
$\tilde{\theta} = 0$. Analogously, if one assumes an equilibrium configuration at high velocities, 
which is typically associated with small, positive angles of attack, the second term on the right hand side of $T_p$ is likely to be positive, thus implying 
a positive~$T_p$ and $\tilde{\theta} = 0$ for positive thrust forces $T$.
Also, simulations that we have performed tend to show that the equilibrium configurations 
associated with $\tilde{\theta} = \pi$ are those requiring an angle of attack belonging to the stall region. 
Then,
stabilizing $\tilde{\theta} = 0$ may ensure that the equilibrium angle of attack does \emph{not belong to the stall region}~\cite[p. 89]{pucciPhd}. 

We have also observed that  
large-constant reference velocities representing a descending phase may be associated with $\tilde{\theta} = 0$, but a negative thrust intensity. Hence, to comply with the  additional constraint $T >0$, one must
stabilize $\tilde{\theta} = \pi$ in these cases. Although this kind of reference velocities are seldom used in practice, 
 remark that the choice of stabilizing either $\tilde{\theta}=0$ or $\tilde{\theta}=\pi$ requires in general close attention:
 stabilizing the former equilibrium does not always ensure a positive thrust at the equilibrium configuration $\vec e_v \equiv 0$.
\subsection{Assumptions}
\begin{hypothesis}
\label{hp:uniquenessEqOr}
There exists a continuous equilibrium orientation $\theta_e(t)$ such that $\tilde{\theta}=0$ and
	\begin{IEEEeqnarray}{RCL}
	  \label{conditionIsolatedTh}
	  |\vec{F}_p(\vec{v}_r(t),\theta_e(t),t)|\,&>& \bar{\eta}, \quad \forall t\,{\in}\ \mathbb{R}_+, \quad \bar{\eta} \in \mathbb{R}_+.
	\end{IEEEeqnarray}
\end{hypothesis}
The above assumption ensures that the control problem is well-posed. In particular,  the continuity of the equilibrium orientation
ensures that no jump of the equilibrium can occur, while the satisfaction of the condition~\eqref{conditionIsolatedTh}
ensures that the equilibrium orientation is differentiable (see Theorem~\ref{th:conditionLS}), so the angular velocity along the reference velocity is defined $\forall t$. 
To avoid non-essential complications, we make the following additional assumption.
\begin{hypothesis}
\label{hp:windDifferentiability}
The aerodynamic coefficients $c_L(\alpha)$ and $c_D(\alpha)$ are twice differentiable functions, 
and their derivatives are bounded $\forall \alpha \in \mathbb{S}^1$.
The wind velocity $\vec{v}_w$ and the reference velocity $\vec{v}_r$, along with their first and second order derivatives,
are  bounded in norm on $\mathbb{R^+}$.
\end{hypothesis}
\subsection{Velocity control}
\label{sec:velocityControl2D}
\begin{proposition}
    \label{velocityControl2D}
    Assume that Assumptions~\ref{hp:uniquenessEqOr}, and~\ref{hp:windDifferentiability} are satisfied.
    Let $k_i>0$, $i = \{1,2,3\}$ and apply the control    
    \begin{IEEEeqnarray}{rCl}
		\label{lawsVGen}
		\IEEEyesnumber
        T &=& \bar{F}_{1}+k_1|F_p|\tilde{v}_1,
        \IEEEyessubnumber
        \label{eq:thrustControlVCe}  \\
        \omega &=& k \left[ k_2|F_p|\tilde{v}_2 +\frac{k_3|F_p| \bar{F}_{p_2}}{(|F_p|+\bar{F}_{p_1})^2}-\frac{\bar{F}_p^T SR^TF_\delta}{|F_p|^2} \right],
        \IEEEyessubnumber \IEEEeqnarraynumspace
        \label{eq:omegaControlVCe}
    \end{IEEEeqnarray}
    to System~\eqref{eq:errorsDynamics} with $\bar{F}_p = R^TF_p$, $\bar{F} = R^TF$, 
    \begin{IEEEeqnarray}{RCL}
	  \label{eq:fAndfpPG}
	  \IEEEyesnumber
	  F &=& mge_1+F_a-m \ddot{x}_r,
	  \IEEEyessubnumber \IEEEeqnarraynumspace \\
	  F_p &=& mge_1+f_p-m \ddot{x}_r, \IEEEyessubnumber \label{FpCV}  \\
	  F_\delta &:=& \partial_{\dot{x}_a}f_p\ddot{x}_a -\partial_{\alpha}f_p\dot{\gamma} -m\dddot{x}_r(t), \IEEEyessubnumber \label{Fdelta}
	\end{IEEEeqnarray}
	\begin{IEEEeqnarray}{RCL}
	  \label{eq:Fafp}
	  \IEEEyesnumber
	  F_a &=&  k_a |\dot{x}_a|[c_L(\alpha)S-c_D(\alpha)I]\dot{x}_a,
	  \IEEEyessubnumber \IEEEeqnarraynumspace \\
	  f_p &=&   k_a |\dot{x}_a|[\bar{c}_{L}(\alpha)S+\bar{c}_{D}(\alpha)]\dot{x}_a, \IEEEyessubnumber
	\end{IEEEeqnarray}
    and $k$ given by:
    \begin{equation}
         k := \bigg(\hspace{0.2cm} 1 \hspace{0.2cm} +\hspace{0.2cm}  \quad \frac{\bar{F}_p^T SR^T}{|F_p|} \partial_{\theta} \left[ \frac{F_p}{|F_p|} \right] \hspace{0.2cm} \bigg)^{-1}.
        \label{eq:kv}
    \end{equation}
    Then,
     \begin{enumerate}
		\item[i)] the control laws~\eqref{lawsVGen} are well defined in a neighborhood of the 
		equilibrium point $(\vec{e}_v,\tilde{\theta}) {=} (0,0)$;
        \item[ii)]  $(\vec{e}_v,\tilde{\theta}) {=} (0,0)$ is a locally asymptotically stable equilibrium point of System~\eqref{eq:errorsDynamics}.
    \end{enumerate}
\end{proposition}
The proof can be found in the Appendix. The interest of System~\eqref{eq:dynamicsVelocityErrorNF}-~\eqref{eq:pGC}
 lies precisely in the expression\footnote{An implementable expression of $k$ is given by
        $\\ k = ( 1  + k_a |\dot{x}_a|^2\bar{F}_{2} (\cos(\alpha{+}\delta)\bar{c}'_D{-}\sin(\alpha{+}\delta)\bar{c}'_L)/|F_p|^2)^{-1}$.} 
        of~$k$. More precisely, 
in view of the result i) of Theorem \ref{th:conditionLS},  the direction of $\vec{F}_p$ is locally
independent of the vehicle's orientation and the nonlinear gain~$k$ is equal to one at the equilibrium configuration. Then, by continuity,  the control law~\eqref{lawsVGen} is well-defined in a neighborhood of the equilibrium point, and stability can be easily proven.

The expression~\eqref{eq:kv} points out that if the vector $\vec{F}_p$  does not depend on the vehicle's orientation $\theta$, e.g. 
the conditions in Lemma~\ref{lemma:conditionLS} are satisfied, then $k\equiv1$. In this case, the control design for 
System~\eqref{eq:dynamicsVelocityErrorNF} can be addressed by adapting the 
methods developed for the class of systems subject to an 
orientation-independent external force. For example, 
\cite{2009_HUA} proposes globally stabilizing controllers for this kind of system. One can then verify that the
velocity control derived by the application of~\cite{2009_HUA} to System~\eqref{eq:dynamicsVelocityErrorNF}
coincides with that given by~\eqref{lawsVGen} with $k \equiv 1$ and $\vec{F}_p$  independent of the orientation $\theta$. 
Also, the control laws~\eqref{lawsVGen} yield a large domain of attraction in this case. 
These facts are stated in the following proposition.
\begin{proposition}
    \label{LargeDomainOfattraction}
    If the aerodynamic coefficients satisfy the condition~\eqref{eq:condition}, then the control laws \eqref{lawsVGen} coincide with the velocity control proposed by \cite{2009_HUA} when applied to system~\eqref{eq:dynamicsVelocityErrorNF}. Consequently, if 
	$ |F_p| > \rho \quad \forall t, \quad \rho > 0,  $  
	and the Assumption~\ref{hp:windDifferentiability} holds,  then the application of the controls~\eqref{lawsVGen} to System~\eqref{eq:dynamicsVelocityErrorNF} renders $(\vec{e}_v,\tilde{\theta}) {=} (0,0)$ an asymptotically stable equilibrium point with domain of attraction 
	equal to $\mathbb{R}^2 \times (-\pi,\pi)$.
\end{proposition}
A consequence
of the above proposition is that when the aerodynamic characteristics are given by 
\eqref{coeffSmallReynolds1} 
any reference velocity is quasi-globally asymptotically stable if
$ |F_p| > \rho \ \forall t$. This latter condition characterizes the set of reference velocities for which the control is not defined. For example, among constant reference velocities and no wind,  the unique reference velocity implying $|F_p| = 0$ is a vertical fall,
a situation rarely met in practice.
\begin{remark}
Once control laws for the asymptotic stabilization of reference velocities are designed, it is straightforward to add integral correction terms for stabilizing reference trajectories. Such an extension can be found at~\cite[p. 83, Sec. 8.3]{pucciPhd}
\end{remark}

\subsection{Control robustification}
\label{controlRobustification}
The control laws~\eqref{lawsVGen} use terms that involve singularities for specific situations. 
To obtain control laws that are well-defined everywhere,          
we first set the nonlinear coefficient $k \equiv 1$  
so that  we do not destroy the local stability property of the above control laws
($k \approx 1$ near the reference trajectory since $\bar{F}_2 \approx 0$). Secondly, we multiply 
the terms $1/(|F_p|+\bar{F}_{p_1})^2$ and $1/|F_p|^2$ by the function $\mu_\tau \in \mathbb{C}^1  :[0,+\infty) \rightarrow [0,1]$, i.e.:
\begin{equation}
	\label{eq:mu2D}
	\mu_\tau(s) =
	\begin{system}
		\sin\left(\tfrac{\pi s^2}{2\tau^2}\right), \quad \text{if } s\leq \tau \\
		1, \quad \quad \quad \quad \text{otherwise}
	\end{system}
\end{equation}
with $\tau > 0$.
This yields the  well-defined expression given by
\begin{IEEEeqnarray}{rCl}
	  \IEEEyesnumber
	\label{eq:inputsVCeR}
 	T &=&  k_1|F_p|\tilde{v}_1 + \bar{F}_{1},
	\label{eq:thrustControlVCeR}
	\IEEEyessubnumber  \\
	\omega &=&  k_2|F_p|\tilde{v}_2 +\mu_\tau(|F_p|+\bar{F}_{p_1}) \tfrac{k_3|F_p| \bar{F}_{p_2}}{(|F_p|{+}\bar{F}_{p_1})^2} \nonumber \\
	&-&\mu_\tau(|F_p|)\tfrac{\bar{F}_p^T SR^TF_{\delta}}{|F_p|^2}.  \IEEEyessubnumber
\end{IEEEeqnarray}
The property: $
	\lim_{s \rightarrow 0} \tfrac{\mu_\tau(s)}{s^2} = \lim_{s \rightarrow 0} \sin\left(\tfrac{\pi s^2}{2\tau^2}\right)s^{-2} = \tfrac{\pi}{2\tau^2},$
implies that the modified control is well-defined everywhere.

\begin{remark}
The control laws~\eqref{lawsVGen}  use the feedforward terms $\ddot{x}_a$, $\dot{\gamma}$ that 
are not always available in practice. Simulations with wing models have shown that neglecting this term when 
its actual value is not too large does not much affect the control performance in terms of ultimate 
tracking errors. 
\end{remark}

\begin{remark}
In practice, the control law~\eqref{eq:inputsVCeR} shall ensure some performance criterion: this may be achieved by relating the control law gains to the chosen  criterion, e.g. the system settling time. Along this direction, the control gain values in the next section have been chosen  using a linearization of the closed loop system at  constant  velocity, and then by  applying an eigenvalue placement technique. The actual value of the eigenvalues may be related to the linearized system settling time. For additional details on this method see~\cite[p. 54]{hua09thesis}.
\end{remark}

\section{Simulations}
\label{sec:simulations}

In this section, we illustrate through a simulation the performance of the proposed approach for the airfoil 
NACA 0021 with the thrust force parallel to the zero-lift-line, e.g. $\delta = 0$.
The  equations of motion are defined by Eqs.~\eqref{eq:dynamics}-\eqref{aerodynamicForce2D} and the aerodynamic coefficients are given by~\eqref{CompleteModel}-\eqref{coefficientsReL}.
The other physical parameters are: 
  $m = 10 \ [Kg]$, $\rho = 1.292 \ \left[ Kg/m^3 \right],$  $\Sigma = 1\ [m^2]$, $k_a = \frac{\rho \Sigma}{2} = 0.6460 \ \left[ Kg/m \right].$ 

We assume here that the control objective is the asymptotic stabilization of a reference velocity and we apply the control laws 
given by~\eqref{eq:inputsVCeR}. 
Other values are used for the calculation of the control laws in order to test the robustness w.r.t. parametric errors. 
They are chosen as follows: $ \hat{m} = 9 \ [Kg]$, $\hat{k}_a =   0.51 \ \left[ Kg/m \right]$, $(c_0,c_1,c_2,c_3) = (20\cdot 10^{-3},0.9,5,0.5)$, 
$\bar{\alpha} = 10^\circ$.  
The feedforward term $F_\delta$ in~\eqref{eq:inputsVCeR} is kept equal to zero, thus providing another element to test the robustness of the controller.
The parameters of the control laws are $k_1 =  0.1529$, $k_2 = 0.0234 $, $k_3 = 6$, $\tau = 80$.

\subsection{From hovering to cruising flight}
The chosen reference velocity represents an hover-to-cruise flight. 
It is composed of: $i)$~an horizontal velocity ramp on the time interval $[0,10) \ [sec]$; $ii)$~cruising with 
constant horizontal velocity of $20 \ [m/sec]$ for $t \geq 10 \ [sec]$. 
Hence, it is given by:  
\begin{IEEEeqnarray}{LLL}
	\label{eq:ramp2D}
	\dot{x}_r(t) =
	\begin{system}
		(0,2t)^T
		\quad \quad  0 \leq &t < 10,     \\
		(0,20)^T
		\quad \quad  \ \ \quad &t \geq  10. \
 	\end{system}
\end{IEEEeqnarray}
Remark that
perfect tracking of this reference velocity is not possible because it  involves discontinuous variations of the 
vehicle orientation (see Section~\ref{subsec:illcond}): using~\eqref{eq:ramp2D}  constitutes another robustness test  of the proposed control strategy. 
The vehicle's initial velocity and attitude are $\dot{x}(0) = (0, 0)^\top$  and $\theta(0)~=~0$ respectively.
No wind is assumed.

\begin{figure}[!t]
  \vspace{0.4cm}
  \centering
  \small{\input{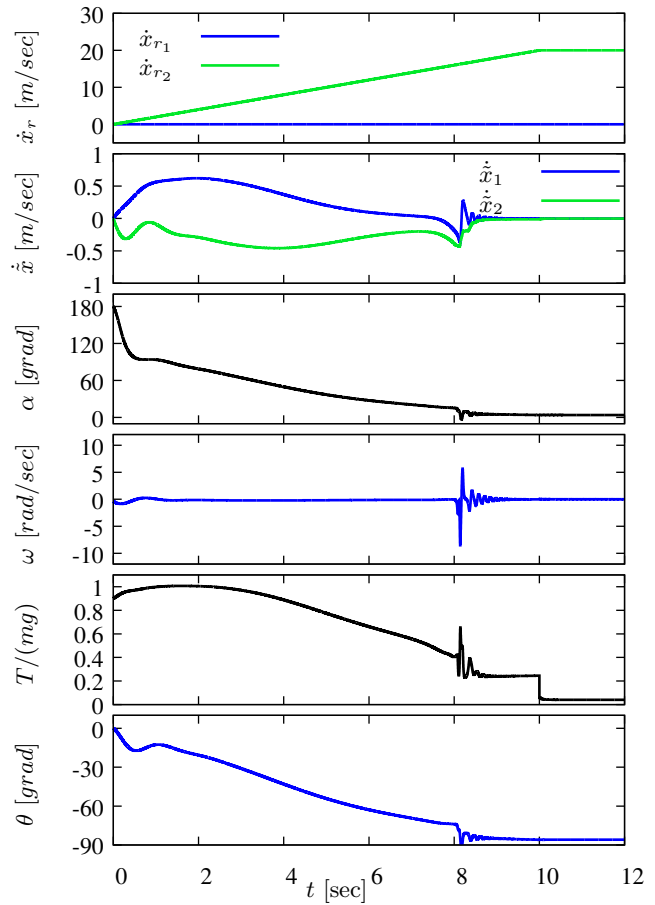}}
  \caption{Simulation of a NACA 0021 profile during hover-to-cruise flight.}
  \label{fig:sim12D}
\end{figure}

Figure~\ref{fig:sim12D} depicts the evolution of the desired reference velocity, the velocity errors, 
the angle of attack, the angular velocity, the thrust-to-weight ratio, and the vehicle orientation. 

At $t = 0$, the vehicle attitude is zero (vertical configuration), and the thrust tends to oppose the body's weight. 
Because of modeling errors and a nonzero reference 
acceleration,
the thrust-to-weight ratio is different from one.
In the interval $(0,10) \ [sec]$, the horizontal velocity of the vehicle increases, the angle of attack decreases,
and the vehicle's orientation tends towards $-90^\circ$ (horizontal configuration).  
At $t = 8$, the equilibrium orientation associated with the reference velocity~\eqref{eq:ramp2D}
\emph{jumps}, and perfect tracking of this reference is not feasible. 
At this time instant, the norm of the vector $\vec{F}_p$ may cross zero. This  generates abrupt variations of the thrust intensity and of the (desired) angular velocity. 
Note that the control value just after the jump depends sensitively upon the constant $\tau$. 
The jump of the equilibrium orientation forbids perfect tracking of the reference velocity:
the velocity errors significantly increase right after the discontinuity occurrence.

\section*{Acknowledgments}
The author is infinitely thankful to Tarek Hamel, Pascal Morin, and Claude Samson for their support, guidance, and help during the
development of this research project.

\section{Conclusions}
Extensions of the vectored--thrust control paradigm~\cite{HUA_2013,naldi2017robust} to the case of aerial vehicles subject to steady aerodynamic forces has been addressed.
Before tackling the control design, we have  studied  the existence and the uniqueness of
the vehicle equilibrium orientation at a reference velocity. 
We have shown that in the case of symmetric shapes, the existence of a vehicle equilibrium orientation can  be asserted independently of the vehicle's shape and reference velocity. 
Concerning the uniqueness, we have presented conditions that ensure either the local or the global uniqueness of the 
equilibrium orientation. The uniqueness of the equilibrium orientation allows us to view the body shape
as a sphere after the application of a specific thrust input. This is what we call \emph{the spherical equivalency}, and  applies to various cases, e.g. flat plates and NACA airfoils flying at low Reynolds numbers: in these case, the application of the proposed thrust input allows one to view  the body with a spherical shape.  Once the transformation is applied, the control design is simplified and  laws for reference velocities and positions can be designed.


Leaving aside the  
adaptations of this work before it is implemented on a device, the 
next step is to extend the analysis presented here to the three-dimensional case, eventually to the case of articulated aircraft~\cite{pucci2017momentum}.
Preliminary achievements in this direction have already been published. For instance, one can show that tboth he global spherical equivalency and the global control design presented in Sec. V and VI can be extended to the case of 3D axis-symmetric aerial vehicles~\cite{phms15}. Furthermore, one can use this result for attempting the control of non-symmetric 3D aircraft~\cite{7039480}. Forthcoming studies may then focus on extending  the in-depth equilibrium analysis presented in the paper to the three-dimensional case.



\bibliographystyle{IEEEtran}
\bibliography{IEEEabrv,./bibliography}

\begin{thebibliography}{10}
\providecommand{\url}[1]{#1}
\csname url@samestyle\endcsname
\providecommand{\newblock}{\relax}
\providecommand{\bibinfo}[2]{#2}
\providecommand{\BIBentrySTDinterwordspacing}{\spaceskip=0pt\relax}
\providecommand{\BIBentryALTinterwordstretchfactor}{4}
\providecommand{\BIBentryALTinterwordspacing}{\spaceskip=\fontdimen2\font plus
\BIBentryALTinterwordstretchfactor\fontdimen3\font minus
  \fontdimen4\font\relax}
\providecommand{\BIBforeignlanguage}[2]{{%
\expandafter\ifx\csname l@#1\endcsname\relax
\typeout{** WARNING: IEEEtran.bst: No hyphenation pattern has been}%
\typeout{** loaded for the language `#1'. Using the pattern for}%
\typeout{** the default language instead.}%
\else
\language=\csname l@#1\endcsname
\fi
#2}}
\providecommand{\BIBdecl}{\relax}
\BIBdecl

\bibitem{2003_STEVENS}
B.~L. Stevens and F.~L. Lewis, \emph{Aircraft Control and Simulation},
  2nd~ed.\hskip 1em plus 0.5em minus 0.4em\relax Wiley-Interscience, 2003.

\bibitem{rdb12}
C.~Roos, C.~D{\"o}ll, and J.-M. Biannic, ``Flight control laws: recent advances
  in the evaluation of their robustness properties,'' \emph{Aerospace Lab},
  vol.~4, p.~., 2012.

\bibitem{2009_KWATNY}
H.~Kwatny, J.-E. Dongmo, B.-C. Chang, G.~Bajpai, C.~Belcastro, and M.~Yasar,
  ``Aircraft accident prevention: Loss-of-control analysis,'' in \emph{AIAA
  Guidance, Navigation, and Control Conference and Exhibit}, 2009.

\bibitem{2010_BELCASTRO}
\BIBentryALTinterwordspacing
C.~Belcastro and J.~Foster, \emph{Aircraft Loss-of-Control Accident
  Analysis}.\hskip 1em plus 0.5em minus 0.4em\relax American Institute of
  Aeronautics and Astronautics, 7 2010. [Online]. Available:
  \url{http://dx.doi.org/10.2514/6.2010-8004}
\BIBentrySTDinterwordspacing

\bibitem{1997_GOMAN}
M.~Goman, G.~Zagainov, and A.~Khramtsovsky, ``Application of bifurcation
  methods to nonlinear flight dynamics problems,'' \emph{Progress in Aerospace
  Sciences}, vol.~33, pp. 539--586, 1997.

\bibitem{1948_PHILLIPS}
W.~Phillips, ``Effect of steady rolling on longitudinal and directional
  stability,'' NACA, Tech. Rep. 1627, 1948.

\bibitem{1974_HACKER}
T.~Hacker and C.~Oprisiu, ``A discussion of the roll-coupling problem,''
  \emph{Progress in Aerospace Sciences}, vol.~15, no.~5, pp. 151--180, 1974.

\bibitem{1998_JAHNKE}
C.~C. Jahnke, ``On the roll-coupling instabilities of high-performance
  aircraft,'' \emph{Philosophical Transactions of the Royal Society of London},
  vol. 356, no. 1745, pp. 2223--2239, 1998, serie A.

\bibitem{2004_STENGEL}
R.~F. Stengel, \emph{Flight Dynamics}.\hskip 1em plus 0.5em minus 0.4em\relax
  Princeton University Press, 2004.

\bibitem{1982_CARROLL}
J.~V. Carroll and R.~K. Mehra, ``Bifurcation analysis of nonlinear aircraft
  dynamics,'' \emph{Journal of Guidance, Control, and Dynamics}, vol.~5, no.~5,
  pp. 529-- 536, 1982.

\bibitem{2004_CUMMINGS}
P.~A. Cummings, ``Continuation methods for qualitative analysis of aircraft
  dynamics,'' NASA, Tech. Rep. CR-2004-213035, 2004.

\bibitem{2002_LOWENBERG}
\BIBentryALTinterwordspacing
M.~Lowenberg, \emph{Bifurcation and Continuation Method}.\hskip 1em plus 0.5em
  minus 0.4em\relax Berlin, Heidelberg: Springer Berlin Heidelberg, 2002, pp.
  89--106. [Online]. Available: \url{http://dx.doi.org/10.1007/3-540-45864-6_6}
\BIBentrySTDinterwordspacing

\bibitem{2000_CHEN}
G.~Chen, J.~L. Moiola, and H.~O. Wang, ``Bifurcation control: theories,
  methods, and applications,'' \emph{International Journal of Bifurcation and
  Chaos}, vol.~10, no.~3, pp. 511--548, 2000.

\bibitem{1991_KWATNY}
H.~G. Kwatny, W.~H. Bennett, and J.~Berg, ``Regulation of relaxed static
  stability aircraft,'' \emph{IEEE Trans. on Automatic Control}, vol.~36,
  no.~11, pp. 1315 -- 1323, 1991.

\bibitem{1990_ABED}
E.~H. Abed and H.~C. Lee, ``Nonlinear stabilization of high angle-of-attack
  flight dynamics using bifurcation control,'' in \emph{American Control
  Conference}, 1990, pp. 2235 -- 2238.

\bibitem{ss80}
S.~N. Singh and A.~Schy, ``Output feedback nonlinear decoupled control
  synthesis and observer design for maneuvering aircraft,'' \emph{International
  Journal of Control}, vol.~31, no.~31, pp. 781--806, 1980.

\bibitem{ws05}
Q.~Wang and R.~Stengel, ``Robust nonlinear flight control of high-performance
  aircraft,'' \emph{IEEE Transactions on Control Systems Technology}, vol.~13,
  no.~1, pp. 15--26, 2005.

\bibitem{mu97}
E.~Muir, ``Robust flight control design challenge problem formulation and
  manual: The high incidence research model (hirm),'' in \emph{Robust Flight
  Control, A Design Challenge (GARTEUR)}, ser. Lecture Notes in Control and
  Information Sciences.\hskip 1em plus 0.5em minus 0.4em\relax Springer Verlag,
  1997, vol. 224, pp. 419--443.

\bibitem{hsm92}
J.~Hauser, S.~Sastry, and G.~Meyer, ``Nonlinear control design for slightly
  non-minimum phase systems: {A}pplication to {V/STOL},'' \emph{Automatica},
  vol.~28, pp. 651--670, 1992.

\bibitem{2002_MARCONI}
L.~Marconi, A.~Isidori, and A.~Serrani, ``Autonomous vertical landing on an
  oscillating platform: an internal-model based approach,'' \emph{Automatica},
  vol.~38, pp. 21--32, 2002.

\bibitem{2003_ISIDORI}
A.~Isidori, L.~Marconi, and A.~Serrani, \emph{Robust autonomous guidance: an
  internal-model based approach.}\hskip 1em plus 0.5em minus 0.4em\relax
  Springer Verlag, 2003.

\bibitem{Bouabdalla05}
S.~Bouabdallah and R.~Siegwart, ``Backstepping and sliding-mode techniques
  applied to an indoor micro quadrotor,'' in \emph{IEEE International
  Conference on Robotics and Automation}, 2005.

\bibitem{Xu08}
R.~Xu and U.~Ozguner, ``Sliding mode control of a class of underactuated
  systems,'' \emph{Automatica}, vol.~44, pp. 233--241, 2008.

\bibitem{kim02}
H.~J. Kim, D.~H. Shim, and S.~Sastry, ``Nonlinear model predictive tracking
  control for rotorcraft-based unmanned aerial vehicles,'' in \emph{American
  Control Conference}, 2002, pp. 3576--3581.

\bibitem{bph07-ifac}
S.~Bertrand, H.~Piet-Lahanier, and T.~Hamel, ``Contractive model predictive
  control of an unmanned aerial vehicle model,'' in \emph{17th IFAC Symp. on
  Automatic Control in Aerospace}, vol.~17, 2007.

\bibitem{hhms13}
M.-D. Hua, T.~Hamel, P.~Morin, and C.~Samson, ``Introduction to feedback
  control of underactuated vtol vehicles,'' \emph{IEEE Control Systems
  Magazine}, pp. 61--75, 2013.

\bibitem{2007_BENOSMAN}
M.~Benosman and K.~Lum, ``Output trajectory tracking for a switched nonlinear
  non-minimum phase system: The vstol aircraft,'' in \emph{IEEE Intl. Conf. on
  Control Applications}, 2007, pp. 262 --269.

\bibitem{2007_FRANK}
A.~Frank, J.~S. McGrew, M.~Valenti, D.~Levine, and J.~P. How, ``Hover,
  transition, and level flight control design for a single-propeller indoor
  airplane,'' in \emph{Guidance, Navigation and Control Conference and Exhibit
  (AIAA)}, 2007, pp. 6318--6336.

\bibitem{1999_OISHI}
M.~Oishi and C.~Tomlin, ``Switched nonlinear control of a vstol aircraft,'' in
  \emph{IEEE Conf. on Decision and Control (CDC)}, 1999, pp. 2685 --2690.

\bibitem{2010_DESBIENS}
A.~L. Desbiens, A.~Asbeck, and M.~Cutkosky, ``Hybrid aerial and scansorial
  robotics,'' in \emph{IEEE Conf. on Robotics and Automation}, 2010, pp. 1--6.

\bibitem{2011_NALDI}
R.~Naldi and L.~Marconi, ``Optimal transition maneuvers for a class of v/stol
  aircraft,'' \emph{Automatica}, vol.~47, no.~5, pp. 870--879, 2011.

\bibitem{2012_CASAU}
P.~Casau, D.~Cabecinhas, and C.~Silvestre, ``Hybrid control strategy for the
  autonomous transition flight of a fixed-wing aircraft,'' \emph{IEEE
  Transactions on control systems technology}, vol.~PP, pp. 1--18, 2012.

\bibitem{phms11CDC}
D.~Pucci, T.~Hamel, P.~Morin, and C.~Samson, ``{Nonlinear Control of PVTOL
  Vehicles subjected to Drag and Lift},'' in \emph{IEEE Conf. on Decision and
  Control}, 2011, pp. 6177--6183.

\bibitem{2012_PUCCI1}
D.~Pucci, ``Flight dynamics and control in relation to stall,'' in
  \emph{American Control Conf. (ACC)}, 2012, pp. 118--124.

\bibitem{2013_PUCCI}
D.~Pucci, T.~Hamel, P.~Morin, and C.~Samson, ``Nonlinear control of aerial
  vehicles subjected to aerodynamic forces,'' in \emph{IEEE Conf. on Decision
  and Control (CDC)}, 2013, pp. 4839 -- 4846.

\bibitem{pucciPhd}
D.~Pucci, ``Towards a unified approach for the control of aerial vehicles,''
  Ph.D. dissertation, Universit\'e de Nice-Sophia Antipolis and ``Sapienza''
  Universita di Roma, 2013.

\bibitem{phms15}
D.~Pucci, T.~Hamel, P.~Morin, and C.~Samson, ``Nonlinear feedback control of
  axisymmetric aerial vehicles,'' \emph{Automatica}, vol.~31, no.~31, pp.
  781--806, 2015.

\bibitem{ghm05}
N.~Guenard, T.~Hamel, and V.~Moreau, ``Dynamic modeling and intuitive control
  strategy for an ''x4-flyer'','' in \emph{International Conference on Control
  and Automation (ICCA2005)}, 2005, pp. 141--146.

\bibitem{hhms09}
M.-D. Hua, T.~Hamel, P.~Morin, and C.~Samson, ``A control approach for
  thrust-propelled underactuated vehicles and its application to {VTOL}
  drones,'' \emph{IEEE Trans. on Automatic Control}, vol. 54(8), pp.
  1837--1853, 2009.

\bibitem{naldi2017robust}
R.~Naldi, M.~Furci, R.~G. Sanfelice, and L.~Marconi, ``Robust global trajectory
  tracking for underactuated vtol aerial vehicles using inner-outer loop
  control paradigms,'' \emph{IEEE Transactions on Automatic Control}, vol.~62,
  no.~1, pp. 97--112, 2017.

\bibitem{kh02}
H.~Khalil, \emph{Nonlinear systems}, 3rd~ed.\hskip 1em plus 0.5em minus
  0.4em\relax Prentice Hall, 2002.

\bibitem{2010_AND}
J.~Anderson, \emph{Fundamentals of Aerodynamics}, 5th~ed.\hskip 1em plus 0.5em
  minus 0.4em\relax McGraw Hill Series in Aeronautical and Aerospace
  Engineering, 2010.

\bibitem{naldi2011optimal}
R.~Naldi and L.~Marconi, ``Optimal transition maneuvers for a class of v/stol
  aircraft,'' \emph{Automatica}, vol.~47, no.~5, pp. 870--879, 2011.

\bibitem{casau2013hybrid}
P.~Casau, D.~Cabecinhas, and C.~Silvestre, ``Hybrid control strategy for the
  autonomous transition flight of a fixed-wing aircraft,'' \emph{IEEE
  Transactions on control systems technology}, vol.~21, no.~6, pp. 2194--2211,
  2013.

\bibitem{7039480}
M.~D. Hua, D.~Pucci, T.~Hamel, P.~Morin, and C.~Samson, ``A novel approach to
  the automatic control of scale model airplanes,'' in \emph{53rd IEEE
  Conference on Decision and Control}, Dec 2014, pp. 805--812.

\bibitem{jung2012development}
Y.~Jung and D.~H. Shim, ``Development and application of controller for
  transition flight of tail-sitter uav,'' \emph{Journal of Intelligent and
  Robotic Systems}, vol.~65, no.~1, pp. 137--152, 2012.

\bibitem{4803783}
W.~E. Green and P.~Y. Oh, ``A hybrid mav for ingress and egress of urban
  environments,'' \emph{IEEE Transactions on Robotics}, vol.~25, no.~2, pp.
  253--263, April 2009.

\bibitem{ccetinsoy2012design}
E.~{\c{C}}etinsoy, S.~Dikyar, C.~Han{\c{c}}er, K.~Oner, E.~Sirimoglu, M.~Unel,
  and M.~Aksit, ``Design and construction of a novel quad tilt-wing uav,''
  \emph{Mechatronics}, vol.~22, no.~6, pp. 723--745, 2012.

\bibitem{muraoka2009quad}
K.~Muraoka, N.~Okada, and D.~Kubo, \emph{Quad Tilt Wing VTOL UAV: Aerodynamic
  Characteristics and Prototype Flight}.\hskip 1em plus 0.5em minus 0.4em\relax
  American Institute of Aeronautics and Astronautics, 2017/05/24 2009.

\bibitem{flores2013transition}
G.~Flores and R.~Lozano, ``Transition flight control of the quad-tilting rotor
  convertible mav,'' in \emph{Unmanned Aircraft Systems (ICUAS), 2013
  International Conference on}.\hskip 1em plus 0.5em minus 0.4em\relax IEEE,
  2013, pp. 789--794.

\bibitem{frank2007hover}
A.~Frank, J.~McGrew, M.~Valenti, D.~Levine, and J.~How, ``Hover, transition,
  and level flight control design for a single-propeller indoor airplane,'' in
  \emph{AIAA Guidance, Navigation and Control Conference and Exhibit}, 2007, p.
  6318.

\bibitem{maqsood2010optimization}
A.~Maqsood and T.~H. Go, ``Optimization of hover-to-cruise transition maneuver
  using variable-incidence wing,'' \emph{Journal of Aircraft}, vol.~47, no.~3,
  pp. 1060--1064, 2010.

\bibitem{itasse2011equilibrium}
M.~Itasse, J.-M. Moschetta, Y.~Ameho, and R.~Carr, ``Equilibrium transition
  study for a hybrid mav,'' \emph{International Journal of Micro Air Vehicles},
  vol.~3, no.~4, pp. 229--245, 2011.

\bibitem{bruntonPhd}
S.~L. Brunton, ``Unsteady aerodynamic models for agile flight at low reynolds
  numbers,'' Ph.D. dissertation, University of Princeton, 2012.

\bibitem{brunton2014state}
S.~L. Brunton, S.~T. Dawson, and C.~W. Rowley, ``State-space model
  identification and feedback control of unsteady aerodynamic forces,''
  \emph{Journal of Fluids and Structures}, vol.~50, pp. 253--270, 2014.

\bibitem{CYBERIAD}
K.~Davis, B.~Kirke, and L.~Lazauskas, ``Material related to the aerodynamics of
  airfoils and lifting surfaces,'' \url{http://www.cyberiad.net/foildata.htm},
  2004, retrieved {\today}.

\bibitem{1996_CHOW}
S.-N. Chow and J.~K. Hale, \emph{Methods of Bifurcation Theory}.\hskip 1em plus
  0.5em minus 0.4em\relax Springer Verlag, 1996.

\bibitem{1990_WIGGINS}
S.~Wiggins, \emph{Introduction to Applied Nonlinear Dynamical Systems and
  Chaos}.\hskip 1em plus 0.5em minus 0.4em\relax Springer Verlag, 1990.

\bibitem{2010_Zhou}
Zhou, M.~Alam, Yang, Guo, and Wood, ``Fluid forces on a very low {R}eynolds
  number airfoil and their prediction,'' \emph{Internation Journal of Heat and
  Fluid Flow}, vol.~21, pp. 329--339, 2011.

\bibitem{tangler2005}
J.~Tangler and J.~D. Kocurek, ``Wind turbine post-stall airfoil performance
  characteristics guidelines for blade-element momentum methods,'' in
  \emph{43rd AIAA Aerospace Sciences Meeting and Exhibit. AIAA}, 2005.

\bibitem{cory2008}
R.~Cory and R.~Tedrake, ``Experiments in fixed-wing uav perching,'' in
  \emph{Proceedings of the AIAA Guidance, Navigation, and Control Conference},
  2008.

\bibitem{moore2012}
J.~Moore and R.~Tedrake, ``Control synthesis and verification for a perching
  uav using lqr-trees,'' in \emph{IEEE Conf. on Decision and Control}, 2012.

\bibitem{2009_HUA}
M.~Hua, T.~Hamel, P.~Morin, and C.~Samson, ``A control approach for
  thrust-propelled underactuated vehicles and its application to vtol drones,''
  \emph{IEEE Transactions on Automatic Control}, vol.~54, pp. 1837--1853, 2009.

\bibitem{hua09thesis}
M.-D. Hua, ``{Contributions to the Automatic Control of Aerial Vehicles},''
  Ph.D. dissertation, Universit{\'e} de Nice-Sophia Antipolis, 2009, available
  at http://hal.archives-ouvertes.fr/tel-00460801/.

\bibitem{HUA_2013}
M.~D. Hua, T.~Hamel, P.~Morin, and C.~Samson, ``Introduction to feedback
  control of underactuated vtol vehicles,'' \emph{IEEE Control Systems
  Magazine}, vol.~33, pp. 61--75, 2013.

\bibitem{pucci2017momentum}
D.~Pucci, S.~Traversaro, and F.~Nori, ``Momentum control of an underactuated
  flying humanoid robot,'' \emph{arXiv:1702.06075}, 2017.

\bibitem{2003_KHALIL}
H.~K. Khalil, \emph{Nonlinear Systems}, 3rd~ed.\hskip 1em plus 0.5em minus
  0.4em\relax Prentice Hall, Upper Saddle River, New Jersey, 2003.

\end{thebibliography}

\section*{Appendix}
\section*{Proof of Lemma~\ref{lemma:dissipativityNotSufficient}}
\label{Plemma:dissipativityNotSufficient}
First, in view of $\vec{F} = (\vec{\imath}_0,\vec{\jmath}_0)F$, $\vec{F}_a = (\vec{\imath}_0,\vec{\jmath}_0)F_a$, 
$\vec{g} = (\vec{\imath}_0,\vec{\jmath}_0)ge_1$, $\vec{\jmath} = (\vec{\imath}_0,\vec{\jmath}_0)Re_2$, 
$\vec{v}_w~=~(\vec{\imath}_0,\vec{\jmath}_0)\dot x_w$,
$\vec{v}_r~=~(\vec{\imath}_0,\vec{\jmath}_0)\dot x_r$, $\vec{a}_r = (\vec{\imath}_0,\vec{\jmath}_0)\ddot x_r$, and 
$\vec{v}_a = (\vec{\imath}_0,\vec{\jmath}_0)\dot x_a$,
the existence of an equilibrium orientation such that \eqref{equilibiumOrinetationCo} 
holds is equivalent to the existence, at any fixed time $t$, of one zero of the following function
\begin{IEEEeqnarray}{RCL}
   f_t(\theta) &:=& F^T(\dot{x}_r(t),\theta,t)R(\theta)e_2,
  \label{eq:wnugeneralp} 
\end{IEEEeqnarray}
where  
\begin{IEEEeqnarray}{RCL}  
   \IEEEyesnumber
    \label{FProofEq}
   F(\dot{x},\theta,t) &=& F_{gr}(t) + F_a(\dot{x}_a,\alpha(\dot{x}_a,\theta)), \IEEEyessubnumber \label{FwithFgrFa} \\
   F_{gr} &:=& mge_1-m \ddot x_r, \IEEEyessubnumber \label{PFgr} \\
   F_a &=& k_a |\dot{x}_a|[c_L(\alpha)S-c_D(\alpha)I]\dot{x}_a, \IEEEyessubnumber \label{PFa} \\
   \dot x_a &=& \dot x - \dot x_w, \IEEEyessubnumber \label{PxaD} \\
   \alpha  &=& \theta -\gamma  +(\pi -\delta) \IEEEyessubnumber \label{Pa} \\
   \gamma &=& \text{atan2}(\dot x_{a_2},\dot x_{a_1}). \IEEEyessubnumber \label{Pg}
\end{IEEEeqnarray}
\noindent
In  coordinates, the aerodynamic force passivity \eqref{dissipativity} 
writes 
\begin{IEEEeqnarray}{r}
  \dot x^T_a F_a \leq 0 \quad \forall (\dot x_a,\alpha). \IEEEeqnarraynumspace \label{Pdissipativity}
\end{IEEEeqnarray}
To show that \eqref{Pdissipativity} does not in general imply the existence of an equilibrium orientation,
it  suffices to find an aerodynamic force satisfying \eqref{Pdissipativity} and such that the function given by
\eqref{eq:wnugeneralp} never crosses zero for some reference and wind velocities at a some time instant. Hence, choose
\begin{equation}
      \label{PcoefficientsPassNoEqui}
	\begin{system}
	  c_{L}(\alpha) &=&  \sin(\alpha) \\
	  c_{D}(\alpha) &=& c_{0} + 1 - \cos(\alpha) > 0, \quad \forall \alpha,
	\end{system}
\end{equation}
with $c_{0} > 0$. It is then straightforward to verify that the aerodynamic force given by \eqref{PFa} with 
the coefficients \eqref{PcoefficientsPassNoEqui} satisfies~\eqref{Pdissipativity}; in addition, note also that 
$c_{L}(0) = c_{L}(\pi) = 0.$
Since the vector $F$ on the
right hand side of~\eqref{eq:wnugeneralp} is evaluated at the reference velocity,
we have to evaluate the quantities \eqref{FProofEq} at  $\dot x_r$. Let us  assume that

{\bf{A1:}} the thrust force is perpendicular to the zero lift direction so that $\delta = \pi/2$;

{\bf{A2:}} there exists a time $\bar t$ such that
 \begin{enumerate}
  \item[i)]  the reference and wind velocities imply 
  $\gamma(\dot x_r(\bar t) {-} \dot x_w(\bar t)){=} \pi/2 
  \ \ \text{ and } \ \ k_a|\dot x_r(\bar t) {-}\dot x_w(\bar t)|^2 {=} 1;   $
  \item[ii)]  the reference acceleration $\ddot x_r(\bar t)$ implies \\
  $F_{gr_1}(\bar t) = 0  \ \ \text{ and } \ \ F_{gr_2}(\bar t) = c_0 + 1. $
 \end{enumerate}
 By evaluating the angle of attack \eqref{Pa}  at the reference velocity with the assumption 
 $\bold{A1}$ and $\bold{A2}$i,
 one  verifies that $\alpha(\bar t) = \theta$. Then, \eqref{eq:wnugeneralp} at $t = \bar t$ becomes 
   $f_{\bar t}(\theta) = [F_{gr_2}(\bar t) {-}c_D(\theta)]\cos(\theta) + [c_L(\theta){-}F_{gr_1}(\bar t)]\sin(\theta). $
In view of the aerodynamic coefficients \eqref{PcoefficientsPassNoEqui} and the assumption $\bold{A}$ii, one has
$f_{\bar t}(\theta) \equiv 1 \ne 0. $
Hence, there exists an aerodynamic force
that satisfies \eqref{Pdissipativity} but for which there does not exist an equilibrium orientation for
some reference and wind velocities at a fixed time instant. 


\section*{Proof of Theorem~\ref{th:existence}}
\label{proof:thExistence}
Recall that the existence of an equilibrium orientation such that~\eqref{equilibiumOrinetationCo} 
holds is equivalent to the existence, at any  time $t$, of one zero of the function $f_t(\theta)$ in 
\eqref{eq:wnugeneralp}.
\subsection*{Proof of the item i)}
Assume that the thrust force is parallel to the zero-lift-line so that $\delta = 0$;
the existence of the equilibrium orientation in the case $\delta = \pi$ can be proven using the same arguments as those below.
Now, in view of Eqs.~\eqref{eq:VaComponents}, 
$\dot x_a = R(\theta) v_a$, $S = R^T(\theta)SR(\theta) $, and of $\delta = 0$, one  verifies that the function $f_t(\theta)$ given by 
\eqref{eq:wnugeneralp} becomes
\begin{IEEEeqnarray}{RCL}
  \label{fThetaSym}
   f_t(\theta) &=&  F^T_{gr}(t)R(\theta)e_2 \nonumber \\
   &-& k_a|\dot x_{rw}(t)|^2[c_L(\alpha_r)\cos(\alpha_r)+c_D(\alpha_r)\sin(\alpha_r)], \nonumber 
\end{IEEEeqnarray}
where $F_{gr}(t)$ is given by \eqref{PFgr} and 
\begin{IEEEeqnarray}{RCL}
  \label{parametersR}
   \alpha_r(\theta,t)  &=& \theta -\gamma_r(t)  +\pi   \label{alphar}  ,\\
   \IEEEyesnumber
   \gamma_r(t) &=& \text{atan2}(\dot x_{rw_2},\dot x_{rw_1}) \IEEEyessubnumber \label{Pgr},\\
   \dot x_{rw}(t) &:=& \dot x_{r}(t) - \dot x_{w}(t).\IEEEyessubnumber \label{xrw}
\end{IEEEeqnarray}
It follows from \eqref{alphar} that at any time $t$ there exists an orientation $\theta_0(t)$ such that 
$\theta = \theta_0(t)$ yields $\alpha_r(t) = 0$, i.e.
\[\theta = \theta_0(t) =\gamma_r(t)-\pi \quad \Rightarrow \quad \alpha_r(t) = 0.\]
Consequently, $\theta = \theta_0(t) + \pi$ yields $\alpha_r(t) = \pi$ and 
$\theta = \theta_0(t) - \pi$ yields $\alpha_r(t) = -\pi$. 
Since it is assumed that the body shape is symmetric, then~\eqref{cLZeroCLPiUZ} holds.
Thus, 
  Eq. \eqref{fThetaSym} yields
\begin{IEEEeqnarray}{RCL}
 \label{propertiesfSym}
 f_t(\theta_0(t) + \pi) &=& f_t(\theta_0(t) - \pi) = -f_t(\theta_0(t))
\end{IEEEeqnarray}
since $e^T_2R^T(\theta_0+\pi)F_{gr}(t)=e^T_2R^T(\theta_0-\pi)F_{gr}(t) = -e^T_2R^T(\theta_0)F_{gr}(t)$.
In view of \eqref{propertiesfSym}, the proof of the existence of (at least) two zeros of the function $f_t(\theta)$ at any fixed time $t$, and thus of
two equilibrium orientations,
is then a direct application of the \emph{intermediate value theorem} since, by assumption, $f_t(\theta)$ is continuous versus $\theta$ ($c_L$ and $c_D$ are continuous) 
and defined $\forall t$ ($\dot{x}_r$ is differentiable). These two zeros, denoted by $\theta_{e_1}(t)$ and $\theta_{e_2}(t)$, 
belong to $\theta_{e_1}(t) \in [\theta_0(t)-\pi, \theta_0(t)]$ and $\theta_{e_2}(t) \in [\theta_0(t), \theta_0(t) + \pi]$. 

\begin{remark}
By looking at the proof of item i), remark that the key assumption is that 
 $c_L(0) = c_L(\pi) = 0$, 
which does not depend on the drag coefficient.
Hence, drag forces have no role in the existence an equilibrium orientation when considering symmetric shapes 
powered by a
thrust force parallel to their axis of symmetry. 
If the thrust force is not parallel to the shape's axis of symmetry, 
one easily shows that the condition $c_L(0) = c_L(\pi) = 0$ is no longer sufficient to ensure the equilibrium orientation existence
for any reference velocity\footnote{Use the same counterexample used to prove Lemma \ref{lemma:dissipativityNotSufficient}.}. 
\end{remark}
\subsection*{Proof of the item ii)}
Under the assumption that the body's shape is bisymmetric, Eqs.~\eqref{LiftPropertiesBSymmetricB} hold, i.e.
	$c_D( \alpha) = c_D( \alpha \pm \pi) \ \forall \alpha$, 
	$c_L( \alpha) = c_L( \alpha \pm \pi) \ \forall \alpha.$
	This property of the aerodynamic coefficients, in view of~\eqref{PFa},  implies
$F_a(\dot{x}_a,\alpha) = F_a(\dot{x}_a,\alpha {\pm} \pi). $ 
Consequently, using the expression
of the angle of attack in~\eqref{Pa}, one verifies that the apparent external force given by \eqref{FwithFgrFa} satisfies
\begin{IEEEeqnarray}{RCL}
  \label{PropFBisymmProof}
  F(\dot{x},\theta,t) = F(\dot{x},\theta \pm \pi,t)  \quad \forall (\dot x,\theta, t). 
\end{IEEEeqnarray} 
In turn, it is straightforward to verify that the function $f_t(\theta)$ given by~\eqref{eq:wnugeneralp} satisfies, at any time $t$, the following 
\begin{IEEEeqnarray}{RCL}
  \label{PropfBisymm2}
  f_t(\theta + \pi) = f_t(\theta - \pi) =-f_t(\theta) \quad \forall \theta. 
\end{IEEEeqnarray} 
Then, analogously to the proof of the  Item 1), the existence of at least two equilibrium orientations  $\theta_{e_1}(t)$ and $\theta_{e_2}(t)$
such that $f_t(\theta_{e_1}(t))= f_t(\theta_{e_2}(t)) = 0$
can be shown by applying the \emph{intermediate value theorem}. 

Observe that Eqs. \eqref{PropfBisymm2} imply that
if $\theta_{e_1}(t)$ is an equilibrium orientation, i.e. $f_t(\theta_{e_1}(t)) = 0 \quad \forall t$, then another equilibrium orientation is given by $\theta_{e_2}(t) = \theta_{e_1}(t) + \pi$.
Now, to show that there always exists an equilibrium orientation ensuring a positive-semi definite thrust intensity, 
from Eq.~\eqref{equilibiumThrustCondition}  observe that the thrust intensity at the equilibrium point is given by
   $T_e = F^T(\dot{x}_r(t),\theta_e(t),t)R(\theta_e(t))e_1.$
Then, it follows from \eqref{PropFBisymmProof} that if the thrust intensity is negative-semi definite at $t$ along an equilibrium orientation, i.e. 
$T_e(\dot{x}_r(t),\theta_{e_1}(t),t) \leq 0$, then it is positive-semi definite at the the equilibrium orientation given by
$\theta_{e_2}(t){=}\theta_{e_1}(t){+}\pi$, i.e. $T_e(\dot{x}_r(t),\theta_{e_1}(t)+\pi,t) \geq 0.$ Hence, one can always build up an equilibrium orientation
$\theta_e(t)$
associated with a positive-semi definite  thrust intensity. 
%

\section*{Proof of Theorem~\ref{th:existenceDeltaDifZeroSymm}}
\label{proof:thExistenceMune0}
First, observe that
if $\sin(\delta) = 0$, then the equilibrium orientation existence follows 
from Theorem \ref{th:existence} since the thrust force is parallel to the zero-lift-direction. Hence assume that 
\begin{IEEEeqnarray}{RCL}
  \label{sinDeltaDiff0}
  \sin(\delta) \ne 0.
\end{IEEEeqnarray} 
Recall that  the equilibrium orientation existence 
is equivalent to the existence, at any fixed time $t$, of one zero of the function $f_t(\theta)$ given by 
\eqref{eq:wnugeneralp}. In view of \eqref{eq:VaComponents}, $\dot x_a = R(\theta) v_a$, and of $S =  R^T(\theta)SR(\theta)$, 
one can verify that \eqref{eq:wnugeneralp} becomes
\begin{IEEEeqnarray}{RCL}
  \label{fThetaBSym}
   f_t(\theta) =  F^T_{gr}(t)R(\theta)e_2 &-& k_a|\dot x_{rw}(t)|^2[c_L(\alpha_r)\cos(\alpha_r+\delta) \nonumber \\
                                         &+&\quad c_D(\alpha_r)\sin(\alpha_r+\delta)], 
  \IEEEeqnarraynumspace
\end{IEEEeqnarray}
where $F_{gr}$ is given by \eqref{PFgr}, 
\begin{IEEEeqnarray}{RCL}
   \alpha_r =\alpha_r(\theta,t) = \theta -\gamma_r(t)  +\pi -\delta, \label{alphar2}
\end{IEEEeqnarray}
$\gamma_r$ by \eqref{Pgr},
and $\dot x_{rw}$  by \eqref{xrw}. 
From Eq. \eqref{fThetaBSym} 
note that if $|\dot x_{rw}(t)| = 0$, 
then there exist at least two zeros for the function $f_t(\theta)$, i.e. at least two equilibrium orientations at the time~$t$. Thus, let us  focus on the following case  
\begin{IEEEeqnarray}{RCL}
  \label{relRefWindDiff0}
  |\dot x_{rw}(t)| \ne 0.
\end{IEEEeqnarray} 
It follows from \eqref{alphar2} that at any fixed time $t$,
there exists an orientation $\theta_0(t) = \gamma_r(t)-\pi+\delta$ such that 
$\theta = \theta_0(t)$ yields $\alpha_r = 0$, so
 $\theta = \theta_0(t) + \pi$ yields $\alpha_r = \pi$. Now, if 
 $f_t(\theta_0(t))f_t(\theta_0(t)+ \pi) \leq 0,$
then there exists a zero for the function $f_t(\theta)$, and this zero belongs to $[\theta_0(t),\theta_0(t)+ \pi]$: in fact, the function $f_t(\theta)$ changes sign on this domain and is continuous versus $\theta$. We are thus interested in the case when the above inequality is not satisfied. 
Hence, 
assume also that
\begin{IEEEeqnarray}{RCL}
  \label{noEquilibriumYet}
  f_t(\theta_0(t))f_t(\theta_0(t)+\pi) > 0.
\end{IEEEeqnarray} 
Given the assumption that the body's shape is symmetric, one has $c_L(0)=c_L(\pi) = 0$. So, in view of \eqref{fThetaBSym}, imposing
\eqref{noEquilibriumYet} divided by $k^2_a|\dot{x}_{rw}(t)|^4\sin(\delta)^2$, which we recall to be assumed different from zero, yields  
\begin{IEEEeqnarray}{RCL}
  \label{noEquilibriumYet1}
  [a_t-c_D(0)][c_D(\pi)-a_t] > 0,
\end{IEEEeqnarray}  
where
  $a_t := \tfrac{F^T_{gr}(t)R(\theta_0(t))e_2}{k_a\sin(\delta)|\dot{x}_{rw}(t)|^2}$. 
Under the assumption that $c_D(0) < c_D(\pi)$,
the inequality \eqref{noEquilibriumYet1} implies that
\begin{IEEEeqnarray}{RCL}
  c_D(0) <a_t< c_D(\pi).   \label{ContrstaintNoEquilibrium}
\end{IEEEeqnarray}  
When the constraint \eqref{ContrstaintNoEquilibrium} is satisfied, the inequality \eqref{noEquilibriumYet} holds and we cannot (yet)
claim the existence of an equilibrium orientation at the time instant $t$. The following shows that when the inequality \eqref{ContrstaintNoEquilibrium} 
is satisfied, the existence of an equilibrium orientation at the time instant $t$ follows from the symmetry of the body's shape provided that the
conditions of Theorem~\ref{th:existenceDeltaDifZeroSymm} hold true. 
Recall that when the inequality \eqref{ContrstaintNoEquilibrium}  is not satisfied, the existence of an equilibrium orientation at the time $t$ follows from the 
fact that 
$f_t(\theta_0(t))f_t(\theta_0(t)+ \pi) \leq 0$.

Now,  under the assumption
that the body's shape is symmetric, we have that Eqs.~\eqref{LiftPropertiesSymmetricB} hold.
Let $\bar \alpha \in \mathbb{R}^+$; then, by using 
    $c_D( \alpha) = c_D( -\alpha)$, $c_L( \alpha)~=~-c_L( -\alpha),$
and \eqref{fThetaBSym},
one verifies that (recall that $\theta = \theta_0(t) \Rightarrow \alpha _r(t) = 0$, so $\theta = \theta_0(t) \pm \bar \alpha \Rightarrow \alpha _r(t) = \pm \bar \alpha$):
\begin{IEEEeqnarray}{RCL}
  \label{noEquilibriumYet2}
  f_t(\theta_0(t)&-& \bar \alpha)f_t(\theta_0(t) + \bar \alpha ) 
             =\Delta_a^2 \sin^2(\delta)-\Lambda_b^2,
  \IEEEeqnarraynumspace  \\
  \label{DeltaLambda}
  \Delta_a=\Delta_a(\bar \alpha)  &:=& [a_t - c_D(\bar \alpha)]\cos(\bar \alpha) + c_L(\bar \alpha)\sin(\bar \alpha), \IEEEeqnarraynumspace  \label{Delta}
\end{IEEEeqnarray} 
$\Lambda_b(\bar \alpha) := [b_t + c_D(\bar \alpha)\cos(\delta)]\sin(\bar \alpha)+ c_L(\bar \alpha)\cos(\bar \alpha)\cos(\delta)$,   $b_t := \tfrac{F^T_{gr}(t)R(\theta_0(t))e_1}{k_a|\dot{x}_{rw}(t)|^2}$.
It follows from Eq. \eqref{noEquilibriumYet2} that if 
\begin{IEEEeqnarray}{RCL}
  \label{Delta1}
  \forall a_t  : c_D(0)  <a_t< c_D(\pi), \exists \bar \alpha_a \in \mathbb{R} \ : \ \Delta_a(\bar \alpha_a) = 0, \IEEEeqnarraynumspace
\end{IEEEeqnarray} 
then there exists a zero for the function $f_t(\theta)$, and this zero belongs to  $[\theta_0(t){-}\bar \alpha_a,\theta_0(t){+}\bar \alpha_a]$ (the function $f_t(\theta)$ would change sign in this domain).
The existence of an $\bar{\alpha}_a$ such that \eqref{Delta1} holds can be deduced by imposing that 
\begin{IEEEeqnarray}{RCL}
  \label{Delta2}
  \forall a_t \ :\ && c_D(0)  <a_t< c_D(\pi), \quad \exists \alpha_0, 
  \alpha_s \in \mathbb{R}, 
  \ : 
  \nonumber \\&& 
  \Delta_a(\alpha_0)\Delta_a(\alpha_s) \leq 0, \IEEEeqnarraynumspace
\end{IEEEeqnarray} 
which implies \eqref{Delta1} with $\bar \alpha_a \in [\alpha_0,\alpha_s]$ since $\Delta_a(\bar \alpha)$ is continuous versus $\bar \alpha$. 
Now, 
in view of \eqref{Delta} note that $\forall a_t : c_D(0)  <a_t< c_D(\pi)$ one has
  $\Delta_a(0) > 0 $.
Still from \eqref{Delta}, note also that $\forall a_t  : c_D(0)  {<}a_t{<}~c_D(\pi)$ one has
\begin{IEEEeqnarray}{RCL}
  \Delta_a(\bar \alpha)  &\leq& c_D(\pi)|\cos(\bar \alpha)| - c_D(\bar \alpha)\cos(\bar \alpha) + c_L(\bar \alpha)\sin(\bar \alpha).\nonumber 
\end{IEEEeqnarray} 
If there exists  $\alpha_s \in (0,90^\circ)$ such that $c_L(\alpha_s) > 0$ and 
\begin{IEEEeqnarray}{RCL}
   c_L(\alpha_s)\sin(\alpha_s){-}[c_D(\alpha_s){-}c_D(\pi)]\cos(\alpha_s) {\leq} 0  \nonumber {\ \Leftrightarrow \ } \text{Cond.~\eqref{conditionMuNe0}} \nonumber \IEEEeqnarraynumspace
\end{IEEEeqnarray} 
then $\Delta_a(\alpha_s) \leq 0$ and \eqref{Delta2} holds with $\alpha_0 = 0$. Consequently, there exists an angle $\bar{\alpha}_a$
such that \eqref{Delta1} is satisfied and, subsequently, an equilibrium orientation $\theta_e(t)$.

\section*{Proof of Lemma~\ref{lemma:conditionLS}}
\label{ProofUnicityGeneric}
 \subsection*{Proof of the item i)}
 The vector $\vec{F}_p$ in~\eqref{eq:FpExt} is independent of the vehicle's orientation $\forall \vec{v}_a$ if and only if the  coefficients $\bar{c}_L$ and $\bar{c}_D$ in~\eqref{eq:fpExt}
 are independent of~$\theta$.
 A necessary and sufficient condition for this independence is that the derivative of $\bar{c}_L$ and $\bar{c}_D$ w.r.t. $\alpha$  equals  zero everywhere. Differentiating~\eqref{clBcdB} w.r.t. $\alpha$ yields
\begin{IEEEeqnarray}{RCL}
	\label{clBcdBD} 
\IEEEyesnumber
	  0 &=&  c'_L(\alpha) -\lambda'\sin(\alpha+\delta)-\lambda \cos(\alpha+\delta), \IEEEyessubnumber \label{clBD}  \\
	  0 &=&  c'_D(\alpha) +\lambda'\cos(\alpha+\delta)-\lambda\sin(\alpha+\delta). \IEEEyessubnumber \label{cdBD}
\end{IEEEeqnarray}
Multiply~\eqref{clBD} by $\cos(\alpha+\delta)$ and~\eqref{cdBD} by $\sin(\alpha+\delta)$.
Summing up the obtained relationships yields~\eqref{eq:pGC}.
Also, multiply~\eqref{clBD} by $\sin(\alpha+\delta)$ and~\eqref{cdBD} by $-\cos(\alpha+\delta)$.
Summing up the obtained relationships yields $\lambda'$. Hence: 
\begin{IEEEeqnarray}{RCL}
	\label{lambdaLambdaP} 
\IEEEyesnumber
	  \lambda  &=&  c'_L(\alpha)\cos(\alpha+\delta) +c'_D(\alpha)\sin(\alpha+\delta), \IEEEyessubnumber \label{lP}  \\
	  \lambda' &=&  c'_L(\alpha)\sin(\alpha+\delta) -c'_D(\alpha)\cos(\alpha+\delta) . \IEEEyessubnumber \label{lPP}
\end{IEEEeqnarray}
Therefore, the vector $\vec{F}_p$ is independent of $\theta$ $\forall \vec{v}_a$  if and only if $\lambda$ and $\lambda'$ are given by~\eqref{lambdaLambdaP}. The function given by~\eqref{lPP}, however, is not always equal to the derivative of~\eqref{lP}. 
By differentiating~\eqref{lP} and by imposing the outcome equal to~\eqref{lPP}, one yields the condition~\eqref{eq:condition}.
 \subsection*{Proof of the item ii)}
One verifies that if the condition~\eqref{eq:condSimpler} is satisfied and $\delta = 0$, 
 substituting~\eqref{eq:pPC}
 in~\eqref{clBcdB} yields $\bar{c}_D(\alpha,\lambda) = \bar{c}_D$, i.e. a constant number, and $\bar{c}_L = 0$. Thus, $\vec{f}_p$ 
 in Eq.\eqref{eq:fpExt}, and $\vec{F}_p$ in Eq.\eqref{eq:FpExt}, are independent of the orientation $\theta$.
 
\section*{Proof of Lemma~\ref{UnicityGeneric}}
Once the velocity errors dynamics are transformed into the form \eqref{eq:dynamicsVelocityErrorNF}, i.e.
	$m\dot{\vec{e}}_v = \vec{F}_p-T_p\vec{\imath}$,
with $\vec F_p$ independent of $\theta$, one verifies that $\vec e_v \equiv 0 $ 
implies
\begin{equation}
  \label{eq:solutionEquG1}
  \begin{system}
	T_p &=& |\vec{F}_p(\vec{v}_r,t)|,  \\
	\vec{\imath}(\theta_e)&=&\tfrac{\vec{F}_p(\vec{v}_r,t)}{|\vec{F}_p(\vec{v}_r,t)|} \Rightarrow \theta_e = \xi_p(t),
  \end{system}
     \hspace{1.0cm}
\end{equation}
\begin{equation}
	\label{eq:solutionEqG2}
	 \begin{system}
		T_p &=& -|\vec{F}_p(\vec{v}_r,t)|,  \\
		\vec{\imath}(\theta_e)&=&-\tfrac{\vec{F}_p(\vec{v}_r,t)}{|\vec{F}_p(\vec{v}_r,t)|} \Rightarrow \theta_e = \xi_p(t)+\pi, 
  \end{system}
\end{equation}
where $\xi_p$ denotes the angle between the vertical direction $\vec{\imath}_0$ and $\vec{F}_p({\vec{v}_r,t})$, i.e. 
$\xi_p = \text{angle}(\vec{\imath_0},\vec{F}_p(\vec{v}_r,t)).$ 
Consequently, at the time instant $t$ one has
 $ \Theta_{\vec{v}_r}(t)=\{ \theta_e(t),\ \ \theta_e(t) + \pi   \}$, 
 if $|\vec{F}_p(\vec{v}_r(t),t)|\ne 0$ and 
 $\Theta_{\vec{v}_r}(t) = \mathbb{S}^1$,
 if $|\vec{F}_p(\vec{v}_r(t),t)|= 0$.  
Then, system~\eqref{eq:dynamicsVelocityErrorNF} has a generically-unique  equilibrium orientation
(see the definition~\ref{def:genericallyUniqueCouple})
if and only if there exists a unique, continuous \emph{bad} reference velocity $\vec{v}_b(t)$ such that 
$|\vec{F}_p(\vec{v}_b(t),t)|= 0 \quad \forall t.$
Now, from Eq.~\eqref{eq:FpExt} observe  that $\vec{F}_p(\vec{v}_b,t)=0 \  \forall t \iff$ 
\begin{IEEEeqnarray}{RCL}
  \label{badSystem}
   \ddot{x}_b= f(\dot{x}_b,t),
\end{IEEEeqnarray}
with $f(\dot{x}_b,t) := ge_1 -\bar{c}|\dot{x}_b-\dot{x}_w|(\bar{c}_LS-\bar{c}_DI)(\dot{x}_b-\dot{x}_w)$, $\bar{c} = k_a/m,$ 
$\vec{v}_w = (\vec{\imath}_0,\vec{\jmath}_0)\dot x_w$, 
$\vec{v}_b = (\vec{\imath}_0,\vec{\jmath}_0)\dot x_b$, and $\vec{a}_b = (\vec{\imath}_0,\vec{\jmath}_0)\ddot x_b$. 
Thus, the problem is to ensure the existence of a unique, continuous solution $\dot{x}_b(t)$ to the 
differential system~\eqref{badSystem}. Without loss of generality, assume that the wind velocity and its 
derivative are bounded, i.e. $\exists c\in\mathbb{R}^+:|\ddot{x}_w|{<}c,|\dot{x}_w|~{<}~c $. 
Then, one  verifies that $f(\dot{x}_b,t)$ is uniformly, locally Lipschitz \cite[p. 90 Lemma 3.2]{2003_KHALIL} on any compact, convex $D_c \subset \mathbb{R}^2$ since
$\exists \delta\in\mathbb{R}^+:|\partial_{\dot{x}_b}f|<\delta, \forall \dot{x}_b \in D_c$. 
So, there exists a unique, continuous solution $\dot{x}_b(t)$ to the differential system~\eqref{badSystem} in $D_c$. 
However, we cannot claim that the solution $\dot{x}_b$ is unique in $\mathbb{R}^2$ since this solution  may 
leave any compact set -- it may tend to infinity in finite time \cite[p. 93 Example 3.3]{2003_KHALIL}. By considering the derivative of the positive-definite function
given by
$V = \tfrac{1}{2}|\dot{x}_b-\dot{x}_w|^2$
with $\bar{c}_D>0$ one  shows that
the solutions to the differential system~\eqref{badSystem} are bounded, so  there exists a convex, 
compact $D_c$ that contains any solution starting at $\dot{x}_b(0) \in \mathbb{R}^2$. Therefore, we deduce that there exists a unique, continuous 
solution to System~\eqref{badSystem} $\forall \dot{x}_b(0) \in \mathbb{R}^2$ and, consequently, to $F(\vec{v} _b,t)=0 \ \forall t$.

\section*{Proof of Lemma~\ref{uniquenessBeautifulModel}}
\label{ProofUniquenessBeautifulModel}
 Assume that the aerodynamic coefficients are independent of the angle $\delta$. Then, the condition~\eqref{eq:condition}
is satisfied for any value of $\delta$ only if
		$c''_D-2c'_L = 0 \ \forall \alpha$, 
		$c''_L+2c'_D = \ \forall \alpha$,
whose  general solution is given by
\begin{equation*}
	\begin{system}
		c_D(\alpha) &=& b_0 +b_1\sin(2\alpha)-b_2\cos(2\alpha), \\
		c_L(\alpha) &=& b_3 + b_1\cos(2\alpha) +b_2\sin(2\alpha),
	\end{system}
\end{equation*}
with $b_j$ denoting constants numbers. When the shape of the body is symmetric, 
the above functions must also satisfy the conditions~\eqref{LiftPropertiesSymmetricB}. 
This implies that $b_1$ and $b_3$ are equal to zero. 
Using the fact that $\cos(2\alpha) {=} 1{-}2\sin^2(\alpha)$, 
one has~\eqref{coeffSmallReynolds1} with $c_0 = b_0 - b_2$ and  $b_2 = c_1$. 

\section*{Proof of Theorem~\ref{th:conditionLS}}
\label{proof:staticFpMagic}
\subsubsection*{Proof of the item $i)$}
To show that the direction of the vector $\vec{F}_p$ is constant w.r.t. the vehicle's orientation at the equilibrium configuration,
we equivalently show that 
\begin{IEEEeqnarray}{rCL}
 \partial_\theta \xi_p \big|_{(\vec{e}_v,\theta)=(0,\theta_e(t))} =0, \label{derXi1}
\end{IEEEeqnarray} 
where
$
 \xi_p := \text{angle}(\vec \imath_0,\vec{F}_p), 
$
and $\vec{F}_p =(\vec{\imath}_0,\vec{\jmath}_0)F_p$. From relations~\eqref{clBcdB},~\eqref{FpTp}, and~\eqref{eq:pGC} one has
\begin{IEEEeqnarray}{RCL}
\IEEEyesnumber
\label{eqs:fpProof}
	  F_p &=& mge_1+f_p-m \ddot{x}_r(t),  \IEEEyessubnumber \label{FpPCP}  \\
	  f_p &=& k_a |\dot{x}_a|[\bar{c}_LS-\bar{c}_DI]\dot{x}_a, \IEEEyessubnumber \label{fpCooI} \\
	  &&\hspace{-1cm}
	  \begin{system}
	  \bar{c}_L &=&  c_L {-}[c'_L\cos(\alpha{+}\delta) 
  {+} c'_D\sin(\alpha{+}\delta)]\sin(\alpha{+}\delta) \\\
	  \bar{c}_D  &=&  c_D {+}[c'_L\cos(\alpha{+}\delta) 
  {+} c'_D\sin(\alpha{+}\delta)]\cos(\alpha{+}\delta),  \IEEEeqnarraynumspace  
	\end{system}   \IEEEyessubnumber \label{cLBcDBP}
	\end{IEEEeqnarray}
where $\vec{a}_r = (\vec{\imath}_0,\vec{\jmath}_0)\ddot{x}_r$. By using~\eqref{FpPCP} and~\eqref{eq:angleOfAttack}, computing the partial derivative of~$\xi_p$ w.r.t. $\theta$  yields
\begin{IEEEeqnarray}{RCL}
\IEEEyesnumber
 \partial_\theta \xi_p &=&-\tfrac{F^T_pS\partial_\theta F_p}{|F_p|^2}=-\tfrac{F^T_pRSR^T\partial_\alpha f_p}{|F_p|^2}, \IEEEyessubnumber 
  \label{derXi2}\\
 \partial_\alpha  f_p &=&k_a |\dot{x}_a|[\bar{c}'_LS-\bar{c}'_DI]\dot{x}_a. \IEEEyessubnumber \label{fpDAL}
\end{IEEEeqnarray} 
Given~\eqref{fpDAL},~\eqref{cLBcDBP}, $\dot x_a = R v_a$, and \eqref{eq:va2}, one verifies that 
\begin{IEEEeqnarray}{rCL}
  \label{propfpB2}
 e^T_2R^T\partial_\alpha f_p \equiv 0. 
\end{IEEEeqnarray} 
Then from Eq. \eqref{derXi2} one obtains
\begin{IEEEeqnarray}{rCL}
  \label{derXi3}
 \partial_\theta \xi_p =-\tfrac{e^T_1R^T\partial_\alpha f_p}{|F_p|^2}e^T_2R^TF_p. 
\end{IEEEeqnarray} 
Now, the transformed System~\eqref{eq:dynamicsVelocityErrorNF} points out that the equilibrium condition 
$\vec{e}_v \equiv 0 $ implies that $\vec{F}_p \cdot \vec{\jmath} \equiv 0 \quad \forall t$. 
This latter condition writes in terms  of coordinates 
\begin{IEEEeqnarray}{rCL}
  \label{equilibriumCondFp}
   e^T_2R^T(\theta_e(t))F_p(\dot{x}_r(t),\theta_e(t),t) = 0 \quad \forall t, 
\end{IEEEeqnarray} 
where $\vec{v}_r = (\vec{\imath}_0,\vec{\jmath}_0)\dot{x}_r$.
By combining~\eqref{equilibriumCondFp} and~\eqref{derXi3}, one shows~\eqref{derXi1} when the condition~\eqref{eq:localUniqueness} is satisfied.
\subsubsection*{Proof of the item $ii)$}
Local uniqueness of an equilibrium orientation is related to 
the equilibrium equation~\eqref{eq:wnugeneralp}.
In particular, if
\begin{IEEEeqnarray}{RCL}
 \partial_\theta f_t(\theta) \big|_{\theta = \theta_e(t)}{=}\partial_\theta   F^T(\dot{x}_r(t),\theta,t)R(\theta)e_2\big|_{\theta {=} \theta_e(t)}  \ne 0, \IEEEeqnarraynumspace
  \label{eq:implicitFunctionCond}
\end{IEEEeqnarray}
then the \emph{implicit function theorem} ensures the existence of a unique differentiable function 
$\theta = \phi(\bar{t})$ satisfying $ f_{\bar{t}}(\theta)=0$ when $\bar{t}$ belongs to a neighborhood 
$\mathcal{I}_{t}$ of $t$ and such that $\theta_e(t)~{=}~\phi(t)$, i.e.
$ f_{\bar{t}}(\phi(\bar{t}))= 0 \ \forall \bar{t} \in  \mathcal{I}_{t}, \ \theta_e(t)~=~\phi(t).$
Hence, the condition~\eqref{eq:implicitFunctionCond} ensures that the equilibrium orientation 
$\theta_e(t)$ is isolated and differentiable at the time instant $t$. 

Using \eqref{eq:VaComponents},
$\partial_\theta F {=} \partial_\theta F_a {=} \partial_\alpha F_a$,  \eqref{FwithFgrFa}, and \eqref{FpPCP}, one gets
\begin{IEEEeqnarray}{RCL}
  \label{propertyDerFAndFp}
  \partial_\theta e^T_2R^TF = -e^T_1R^TF_p.
\end{IEEEeqnarray}
Now, it is a simple matter to verify that
\begin{IEEEeqnarray}{RCL}
  \label{FB2FpB2}
  e^T_2R^TF  \equiv e^T_2R^TF_p.   
\end{IEEEeqnarray}
Also, recall that at the equilibrium one has:
 \[ e^T_2R^TF\big|_{(\dot x,\theta) \equiv (\dot x_r,\theta_e)}\equiv0,\] 
therefore, in view of  \eqref{FB2FpB2}, at the equilibrium one has 
\begin{IEEEeqnarray}{RCL}
  e^T_2R^TF_p\big|_{(\dot x,\theta) \equiv (\dot x_r,\theta_e)}&\equiv&0, \label{Fp2BEq} 
\end{IEEEeqnarray}
i.e. the second component of the vector $R^TF_p$ is equal to zero at the equilibrium point. Now, in view of 
$|F_p|~=~|R^TF_p|$,
and of the assumption that the vector $F_p$ is different from zero at the equilibrium point, one has that 
the first component of the vector $R^TF_p$ is necessarily different from zero at the equilibrium point, i.e.
  $e^T_1R^TF_p\big|_{(\dot x,\theta) \equiv (\dot x_r,\theta_e)} \ne 0. $
This in turn implies, via \eqref{propertyDerFAndFp}, that the condition \eqref{eq:implicitFunctionCond} is satisfied. Consequently, the equilibrium orientation $\theta_e(t)$ is isolated and differentiable at the time instant $t$.

\subsubsection*{Proof of the item $iii)$}

Using  
$\vec{e}_v:=(\vec{\imath}_0,\vec{\jmath}_0)\dot{\tilde{x}}$ and System~\eqref{eq:dynamicsVelocityErrorNF} yields the following tracking error dynamics 
\begin{IEEEeqnarray}{RCL}
\IEEEyesnumber
	\label{eq:dynamicsVelocityErrorNFP}
	m\ddot{\tilde{x}} &=& F_p(\dot{x},\theta,t)-T_p R(\theta)e_1, \IEEEyessubnumber 
	\\ \dot{\tilde{\theta}} &=& \omega - \dot{\theta}_e(t), \IEEEyessubnumber
	\label{eq:dynamicsVelocityErrorGUCP} 
\end{IEEEeqnarray}
where $\tilde{\theta} := \theta - \theta_e(t)$. In view of the item ii), 
the equilibrium orientation $\theta_e(t)$ is differentiable, so  the Eq.~\eqref{eq:dynamicsVelocityErrorGUCP} 
is well-conditioned. Now,
take $(T_p,\omega)$ as control inputs and $(m{\dot{\tilde{x}}},\tilde{\theta})$ as state variables. 
Observe that
$ \partial_\theta F_p = \tfrac{F_p}{|F_p|} \partial_\theta  |F_p| + |F_p| \partial_\theta \left[ \tfrac{F_p}{|F_p|}\right],$
and that $F_p/|F_p|=\pm Re_1$ and $T_p = \pm |F_p|$ at the equilibrium $(m{\dot{\tilde{x}}},\tilde{\theta}) {=} (0,0)$. 
In view of 
the item i), 
one shows that the state and control matrices associated with 
the linearization of System~\eqref{eq:dynamicsVelocityErrorNFP} 
are given by
\begin{IEEEeqnarray}{RCL}
	A & = & 
		\begin{pmatrix} 
			 \tfrac{\partial_{\dot{x}}F_p}{m}(\dot{x}_r,\theta_e,t)  & a(t)R(\theta_e) e_1 +b(t)R(\theta_e) e_2 \\ 
			0_{1\times2}  &  0 
		\end{pmatrix}, \nonumber \\
	B  &=&  
		\begin{pmatrix} 
			-R(\theta_e)e_1 &  0_{2\times1} \\ 
			0  &  1 
		\end{pmatrix}, \nonumber
\end{IEEEeqnarray}
where $0_{n\times m} \in \mathbb{R}^{n\times m}$ denotes a matrix of zeros, 
$a(t) := \pm \partial_{\theta}| F_p| (\dot{x}_r(t),\theta_e(t),t)$, and 
$b(t) := \pm|F_p(\dot{x}_r(t),\theta_e(t),t)|$. When the condition~\eqref{eq:localUniqueness} holds, one has $b(t) \ne 0 \ \forall t$ and 
it is a simple matter to verify that the 
matrix
$
	\begin{pmatrix} 
		B & AB {-} \dot{B}
	\end{pmatrix}$
is of full rank, which  implies the controllability of~\eqref{eq:dynamicsVelocityErrorNF}.


\section*{Proof of Proposition~\ref{velocityControl2D}}
\label{proof:velocityControl2D}

By decomposing the velocity errors in the body frame basis, i.e. 
$\vec e_v =(\vec \imath,\vec \jmath) \tilde v$, from System \eqref{eq:errorsDynamics}
 one obtains
\begin{IEEEeqnarray}{rCL}
  \label{dynamicsErrorsVP}
  m\dot{\tilde v} &=& -m\omega S \tilde v -T_p e_1 + \bar F_p ,
  \quad 
  \dot{\theta} = \omega,
\end{IEEEeqnarray}
with $\bar F_p = R^T F_p$, and $F_p$ given by \eqref{eqs:fpProof}.
Recall that $\tilde{\theta} \in (-\pi,\pi]$ denotes the angle between the two vectors $e_1$ and $\bar{F}_p$ 
so that 
$|F_p|\cos(\tilde{\theta}) = \bar{F}_{p_1}$, and that the control objective is the asymptotic stabilization of 
$\tilde{\theta}$ to zero. Then, we consider the candidate Lyapunov function defined by:
	\begin{equation}
		V= \frac{m}{2} |\tilde{v}|^2 +\frac{1}{k_2} \bigg(1-\frac{\bar{F}_{p_1}}{|F_p|}\bigg).
		\label{eq:lyFunNC}
	\end{equation}
Now, to compute $\dot{V}$, first verify that for any $x\in\mathbb{R}^2$ one has
\begin{IEEEeqnarray}{RCL}
 |x|^2I+Sxx^\top S = xx^\top. \nonumber
\end{IEEEeqnarray} 
By using $\dot{R}=\omega SR$, $RS=SR$, $S^2 = -I$, and the above identity,  one can verify that
	\begin{IEEEeqnarray}{RCL}
		\frac{d}{dt} \bigg(1-\tfrac{\bar{F}_{p_1}}{|F_p|}\bigg) 
		&=&\tfrac{e^\top_1}{|F_p|}\left[-\omega S \bar{F}_p+R^\top\left(I-\tfrac{F_pF^{\top}_p}{|F_p|^2}\right)\dot{F}_p \right]  \nonumber \\ &=&
	-\tfrac{\bar{F}_{p_2}}{|F_p|} \left(\omega + \tfrac{F_p^T S\dot{F}_p}{|F_p|^2} \right).
		\label{derOrientationError}
	\end{IEEEeqnarray}
In view of \eqref{eqs:fpProof},  $F_\delta$ as~\eqref{Fdelta}, 
$\partial_\alpha f_p = \partial_\theta f_p = \partial_\theta F_p$, and $\alpha = \theta-\gamma +\pi-\delta$, the term $\dot{F}_p$ in the right hand side of the above equation becomes
\begin{IEEEeqnarray}{RCL}
\label{FpDot}
\dot{F}_p &=& \partial_{\dot{x}_a}f_p \ \ddot{x}_a + \partial_{\alpha}f_p \ \dot{\alpha}  -m\dddot x_r \nonumber \\ &=& \partial_{\dot{x}_a}f_p \ \ddot{x}_a -\partial_{\alpha}f_p \ \dot{\gamma} + \partial_{\alpha}f_p \ \omega  -m\dddot x_r= F_\delta+ \partial_{\alpha}f_p \ \omega \nonumber \\ &=& F_\delta+ \partial_{\theta}F_p \ \omega.
\end{IEEEeqnarray}
Observe also that 
\begin{IEEEeqnarray}{RCL}
\label{FpTheta}
\partial_\theta F_p &=& 
\partial_\theta 
\left(|F_p| \frac{F_p}{|F_p|} \right)= 
 \tfrac{F_p}{|F_p|} \partial_\theta  |F_p| + |F_p| \partial_\theta \left[ \tfrac{F_p}{|F_p|}\right].
\end{IEEEeqnarray}
In light of \eqref{derOrientationError},~\eqref{FpDot},  and~\eqref{FpTheta} 
the function $\dot{V}$ along the solutions 
of System~\eqref{dynamicsErrorsVP} becomes
\begin{IEEEeqnarray}{rcl}
	\label{dotVPvelocityControl}
	\dot{V} &=& \tilde{v}_1(\bar{F}_{p_1}-T_p)  
 	\\ &{-}&\
	\tfrac{\bar{F}_{p_2}}{k_2|F_p|} \left[\left(1{+}\tfrac{F_p^TS}{|F_p|}\partial_{\theta}\left(\tfrac{F_p}{|F_p|}\right) \right)\omega {+} 
	\tfrac{F_p^T SF_{\delta}}{|F_p|^2} {-}k_2|F_p|\tilde{v}_2\right]. 
	 \nonumber
\end{IEEEeqnarray}
The term multiplying $\omega$ 
is equal to one, and thus different from zero, at to the equilibrium point (see Theorem~\ref{th:conditionLS}). 

The application of the control laws $(T,\omega)$ given by \eqref{lawsVGen} thus yields in a neighborhood of the equilibrium point
\begin{IEEEeqnarray}{rCL}
	\dot{V} &{=}&
		{-}k_1|F_p|\tilde{v}_1^2 {-}\tfrac{k_3}{k_2} \tan^2\left(\tilde{\theta}/{2}\right),
		\label{eq:endProofControl1NC}  \IEEEeqnarraynumspace
\end{IEEEeqnarray}
because 
$\quad \tan^2(\tilde{\theta} / 2)=\bar{F}_{p_2}^2/(|F_p|+\bar{F}_{p_1})^2.$
Since $\dot{V}$ is negative semi-definite, the velocity error term $\tilde{v}$ is locally bounded. 
The next step of the proof consists in showing the uniform continuity of $\dot{V}$ along every system solution and, 
using Barbalat's lemma, one deduces the convergence of $\tilde{v}$ and $\tilde{\theta}$ to zero. 
In order to prove the mentioned uniform continuity of $\dot{V}$, it suffices to show that $\ddot{V}$ is bounded. Note that in view of Assumption~\ref{hp:uniquenessEqOr}, the vector $F_p$ is different from zero in an open neighborhood of $(\tilde{v},\tilde{\theta}) = (0,0)$. 
Consequently, it is a simple matter to verify that there exists an open neighborhood of 
$(\tilde{v},\tilde{\theta}) = (0,0)$ in which $\ddot{V}$ is bounded. As a consequence, in view of~\eqref{eq:endProofControl1NC}, there exists an open neighborhood of 
$(\tilde{v},\tilde{\theta}) = (0,0)$ such that for any initial condition in it, 
$\tilde{v}_1$ and $\tilde{\theta}$ converge to zero. 
Now, in order to show that $\tilde{v}_2$ tends to zero, observe that 
\begin{equation}
	\label{eq:convV2}
 	\frac{d}{dt}\tfrac{\bar{F}_{p_2}}{|F_p|} = -k_2\bar{F}_{p_1}\tilde{v}_2 +k_3\tfrac{\bar{F}_{p_1}\bar{F}_{p_2}}{(|F_p|+\bar{F}_{p_1})^2},
\end{equation}
where the control inputs $(T,\omega)$ were chosen as \eqref{lawsVGen}.
By applying Barbalat's Lemma, one verifies the uniform continuity of~\eqref{eq:convV2} in a neighborhood of $(\tilde{v},\tilde{\theta}) = (0,0)$. Then, the right hand side of \eqref{eq:convV2} tends to zero. Since \[\bar{F}_{p_2} \rightarrow 0, \quad (|F_p|+\bar{F}_{p_1})^2 > 0, \quad \bar{F}_{p_1} \rightarrow |F_p| > 0\] in a neighborhood of $(\tilde{v},\tilde{\theta}) = (0,0)$, then there exists an open neighborhood of $(\tilde{v},\tilde{\theta}) = (0,0)$ such that for any initial condition in it, $\tilde{v}_2$ necessarily tends to zero. As for the stability of the equilibrium $(\tilde{v},\tilde{\theta}) = (0,0)$, it is a consequence of relations~\eqref{eq:lyFunNC} and~\eqref{eq:endProofControl1NC}.


%
%
%

%
%

%

\end{document}